\documentclass[11pt]{article}
\usepackage{epstopdf}
\usepackage[a4paper,hmarginratio=1:1,vmarginratio=2:3,totalwidth=15.3cm,totalheight=22.7cm]{geometry}
\usepackage{bm,epstopdf,epsfig,amsmath,amssymb,amsfonts,colordvi,latexsym,comment,cancel,
verbatim,extarrows,slashed}
\usepackage{graphicx,graphics}
\usepackage[font=md,captionskip=8pt]{subfig}
\usepackage[usenames,dvipsnames]{color}

\usepackage{setspace}
\usepackage[utf8]{inputenc}
\allowdisplaybreaks

\newcommand{\barloga}{\overline\ln V_p}
\newcommand{\barlogb}{\overline\ln V_m}

\newcommand{\cO}{{\cal O}}

\newcommand{\ra}{\rightarrow}
\newcommand{\be}{\begin{equation}}
\newcommand{\ee}{\end{equation}}
\newcommand{\bea}{\begin{eqnarray}}
\newcommand{\eea}{\end{eqnarray}}

\newcommand{\baa}{\begin{array}}
\newcommand{\eaa}{\end{array}}
\long\def\symbolfootnote[#1]#2{\begingroup
\def\thefootnote{\fnsymbol{footnote}}\footnote[#1]{#2}\endgroup}
\newcommand{\laf}{\lambda_\phi}
\newcommand{\lam}{\lambda_m}
\newcommand{\las}{\lambda_\sigma}

\begin{document} 
\begin{flushright}
CERN-PH-TH-2016-186\\
%\today
\end{flushright}

\bigskip\medskip

\thispagestyle{empty}

\vspace{2cm}

\begin{center}
  {\Large {\bf Two-loop scale-invariant scalar  potential}\\

 \bigskip
 {\bf  and quantum effective operators}}
% \\

\vspace{1.cm}

 {\bf D. M. Ghilencea}$^{\,a,b}$, \,\,  {\bf Z. Lalak}$^{\,c}$ and \,\, {\bf  P. Olszewski}$^{\,c}$ 
\symbolfootnote[1]{E-mail: dumitru.ghilencea@cern.ch, 
pawel.olszewski@fuw.edu.pl, zygmunt.lalak@fuw.edu.pl}

\bigskip

$^a$ {\small  Theory Division, CERN, 1211 Geneva 23, Switzerland}

$^b$ {\small Theoretical Physics Department, National Institute of Physics}

{\small and Nuclear\, Engineering \, (IFIN-HH)\, Bucharest\, 077125, Romania}

$^c$ {\small Institute of Theoretical Physics, Faculty of Physics, University of Warsaw}

{\small  ul. Pasteura 5, 02-093 Warsaw, Poland}

\end{center}

\begin{abstract}
\noindent
Spontaneous breaking of  quantum scale invariance may provide a solution to 
the hierarchy and cosmological constant problems.  In a 
scale-invariant regularization, we compute the two-loop potential of  a higgs-like 
scalar  $\phi$  in theories in which  scale symmetry is broken only 
spontaneously by the dilaton ($\sigma$).
Its  vev $\langle\sigma\rangle$ generates the DR subtraction scale ($\mu\sim\langle\sigma\rangle$), 
which  avoids the explicit scale symmetry breaking by traditional regularizations (where 
$\mu$=fixed scale).
The  two-loop potential contains  effective operators of  non-polynomial 
nature as well as new corrections, beyond those obtained with  explicit breaking  ($\mu$=fixed scale).
These operators  have the form: $\phi^6/\sigma^2$, $\phi^8/\sigma^4$, etc, 
which generate an  infinite series of higher dimensional  polynomial  operators 
upon  expansion about  $\langle\sigma\rangle\gg \langle\phi\rangle$, where
 such hierarchy is arranged by {\it one}  initial, classical tuning.
 These operators  emerge at the quantum  level from  evanescent interactions ($\propto\epsilon$)  
between $\sigma$ and $\phi$ that  vanish in $d=4$ but are demanded by classical scale 
invariance in $d=4-2\epsilon$. The Callan-Symanzik equation of the two-loop
 potential is respected  and the two-loop  beta functions of the couplings differ 
from those of the same theory regularized with $\mu=$fixed  scale.
Therefore the  running of the couplings enables one to distinguish between spontaneous and 
explicit scale symmetry breaking.
\end{abstract}

\newpage

\section{Introduction}

Theories with scale symmetry  \cite{W} may provide a solution to  the
 hierarchy and cosmological constant problems.
But scale symmetry is not a symmetry of the real world, therefore it must be broken.
In this work we discuss theories with scale invariance at the classical
and quantum level that is broken {\it  only spontaneously}. 
This is important since in a classical scale invariant
theory, quantum calculations usually  break this symmetry {\it explicitly} due to the presence
of the  subtraction (renormalization) {\it scale}  ($\mu$).
This scale is introduced to regularize the loop integrals,
regardless of the  regularization method: dimensional regularization  (DR),  
Pauli-Villars, etc, and its simple  presence breaks explicitly this symmetry.

It is known however how to avoid this problem by using a subtraction scale that 
is generated {\it spontaneously},  as the vacuum expectation value (vev) of  a scalar 
field $\sigma$ \cite{Englert}. This field is the Goldstone mode of scale 
symmetry (dilaton) and then $\mu=z \langle\sigma\rangle$, where $z$ is a dimensionless 
parameter.  But before (spontaneous) scale symmetry breaking, with a field-dependent 
subtraction  {\it function} $\mu(\sigma)= z \sigma$,  there is no scale in the theory.
One can use this idea to compute quantum corrections  to 
the scalar potential of a theory with a higgs-like scalar $\phi$ 
and dilaton $\sigma$  and obtain a scale invariant result  at one-loop 
\cite{S1,S2,S3,dmg,P}  with a flat direction and spontaneous scale symmetry breaking.
Although the result is scale invariant at the quantum level, the couplings still run 
with the momentum scale \cite{S3,dmg,tamarit}\footnote{After 
spontaneous breaking of scale symmetry
 $\langle\sigma\rangle\not=0$, the subtraction scale $\mu(\langle\sigma\rangle)$
 and all other masses/vev's of the theory are generated, proportional to $\langle\sigma\rangle$.}.

To illustrate some of these ideas, consider a scale invariant theory in $d=4$ 
\medskip
\bea
L=
\frac{1}{2}\partial_\mu \phi\partial^\mu\phi
+\frac{1}{2}\partial_\mu\sigma\partial^\mu\sigma
-V(\phi,\sigma)
\eea

\medskip\noindent
where $\phi$ is a higgs-like scalar and $\sigma$ is a dilaton.
In such a theory $V$  has a form
%\medskip
\bea\label{form}
V(\phi,\sigma)=\sigma^4\, W(\phi/\sigma)
\eea
%\medskip\noindent
In this paper we  assume 
that we have  spontaneous breaking of this symmetry, hence  $\langle\sigma\rangle\!\not=\!0$.
We do not detail how $\sigma$ acquires a vev (expected to be large  
$\langle\sigma\rangle\sim M_\text{Planck}$)
 but search for solutions with  $\langle\sigma\rangle\!\not=\!0$.
Then the two minimum conditions $\partial V/\partial\phi\!=\!\partial V/\partial\sigma\!=\!0$
become
%\medskip
\bea\label{W}
W'(x_0)=W(x_0)=0,
\qquad 
x_0\equiv\frac{\langle\phi\rangle}{\langle\sigma\rangle};\,
\quad \langle\sigma\rangle, \langle\phi\rangle\not=0.
\eea 

\medskip\noindent
At a given order $n$ in perturbation theory, 
one condition, say $W^\prime(x_0)=0$,
fixes the ratio $x_0\equiv \langle\phi\rangle/\langle\sigma\rangle$ in terms of the (dimensionless)
couplings of the theory. The second condition, $W(x_0)=0$, leads to
vanishing vacuum energy $V(\langle\phi\rangle,\langle\sigma\rangle)=0$ and 
fixes a relation among
the couplings, corrected to  that order ($n$)  in perturbation theory from its version in 
the lower perturbation  order ($n-1$). If these two equations have
a solution $x_0$, then the system has a flat direction (Goldstone) in the plane
$(\phi,\sigma)$ with $\phi/\sigma=x_0$. Then a 
massless state exists (dilaton) at this order.
This is true  {\it provided that} quantum  corrections
do not break {\it explicitly}  the scale symmetry (otherwise,  eq.(\ref{form}) is not valid
 due to the presence of  the ``usual'' DR scale $\mu$). With a scale invariant
regularization, it is  possible to keep these properties
($V=0$,  a flat direction, etc) and study spontaneously broken  quantum scale invariance.

Why is this interesting? One reason is that this answers the question of Bardeen
\cite{Bardeen} on  the mass hierarchy.
The Standard Model (SM) with a vanishing classical higgs mass term is scale invariant
and there is no mass hierarchy  (ignoring gravity, as here\footnote{For related applications 
that include gravity, see for example \cite{FHR,S4,O,KA}.}).
If quantum calculations preserve this symmetry, via a scale invariant regularization,
 one can avoid a  hierarchy problem and the fine-tuning of the higgs self-coupling and
 keep it light relative to the  high scale (physical mass of a new state)
generated by $\langle\sigma\rangle\not=0$.
 One can arrange that  $x_0=\langle\phi\rangle/\langle\sigma\rangle\!\ll\! 1$ by 
a {\it single}  classical tuning of the (ratio of the) couplings of the 
theory \cite{GGR}.  The hierarchy 
$m_\text{higgs}^2\sim \langle\phi\rangle^2\!\ll\! \langle\sigma\rangle^2$
 is maintained  at one-loop  \cite{S1,S2,S3,dmg,P,GGR,FK} and probably
 beyond it, due to the spontaneous-only scale symmetry  breaking.
The only difference from the  usual SM is the presence of a massless
 dilaton in addition to the SM spectrum. Also,  the solution $x_0$ is related 
to the (minimum) condition $V=0$. This suggests that in spontaneously broken
 quantum scale invariant theories any fine tuning is   related to vacuum energy 
tuning at the same order of perturbation.

With this motivation, in this paper  we extend the above  results.
 We consider a classically scale invariant theory of
 $\phi$ and $\sigma$ and   compute at  two-loop the scalar potential and the 
running of the couplings,  in a   scale invariant regularization.
We find that starting from two loops, the running of the couplings differs from that in
 the same  theory of $\phi$, $\sigma$ regularised with $\mu\!=$constant.
We  show that effective {\it non-polynomial} operators like
 $\phi^6/\sigma^2$, $\phi^8/\sigma^4$, are generated as two-loop counterterms.
 If expanded about the ground state, these operators 
generate an infinite series of polynomial terms,  showing the non-renormalizability 
of the theory. The Callan-Symanzik equation of the potential is  verified at two loops. 
The results are useful for phenomenology, {\it e.g.} to study a scale invariant version of the 
SM (+dilaton).

\section{One-loop potential}

We first review the one-loop  potential \cite{dmg,P}. Consider the classical potential\footnote{
In principle one can also include higher dimensional 
terms like $\phi^6/\sigma^2$, $\phi^8/\sigma^4$, etc, ($\langle\sigma\rangle\not=0$), see later.}
%\medskip
\bea
V = \frac{\lambda_\phi}{4!} \phi^4 + \frac{\lambda_m}{4} \phi^2 \sigma^2 
+ \frac{\lambda_\sigma}{4!} \sigma^4.
\eea
%
%\medskip\noindent
Spontaneous scale symmetry breaking $\langle\sigma\rangle\not=0$ 
demands two conditions (eq.(\ref{W}))  be met:
\bea\label{minimum}
9 \lambda_m^2=\lambda_\phi\lambda_\sigma+\textsf{loops},\,\,(\lambda_m<0),
\qquad\text{and}\qquad
x_0^2\equiv \frac{\langle\phi\rangle^2}{\langle\sigma\rangle^2}=
-\frac{3 \lambda_m}{\lambda_\phi}+\textsf{loops.}
\eea
A massless (Goldstone) state exists corresponding  to a flat 
direction $\phi=x_0\,\sigma$ with $V_\textsf{min}=0$. 
With $\phi$ being higgs-like, scale symmetry breaking implies electroweak symmetry breaking.

To compute quantum corrections in $d=4-2\epsilon$, the scalar potential is modified 
to  $\tilde V=\mu^{2\epsilon}V$ to ensure dimensionless quartic couplings, with $\mu$ the 
``usual'' DR subtraction  scale. General 
principles\footnote{They demand  quantum interactions between  $\phi$ and $\sigma$ vanish 
in their classically decoupling limit $\lambda_m\!=\!0$.} suggest 
that the subtraction {\it function} $\mu(\sigma)$ depend on the dilaton only \cite{dmg} and
generate the subtraction scale $\mu(\langle\sigma\rangle)$
after spontaneous scale symmetry breaking;
 $\mu(\sigma)$ is then  identified  on dimensional grounds (using $[\mu]=1$, $[\sigma]=(d-2)/2$).
Then the  scale invariant potential in $d=4-2\epsilon$ and $\mu(\sigma)$ become
\bea\label{tildeV}
\tilde V(\phi,\sigma)=\mu(\sigma)^{2\epsilon} V(\phi,\sigma),
\qquad \mu(\sigma)=z\,\sigma^{1/(1-\epsilon)},
\eea
 where  $z$ is   an arbitrary dimensionless  parameter\footnote{The parameter $z$ plays a 
special role in the Callan Symanzik equation,   see later.}. 
The one-loop result is
\bea
V_1&=&\tilde V -
\frac{i}{2}\,
\,\int \frac{d^d p}{(2\pi)^d}
\,{\rm Tr}\ln \big[ p^2-\tilde V_{\alpha\beta}+i\varepsilon\big]
\eea

\medskip\noindent
Here  $\tilde V_{ij}=\partial^2 \tilde V/\partial s_i \partial s_j$, \, ($i,j=1,2$),
 $s_1=\phi$, $s_2=\sigma$ and similar for   $V_{ij}=\partial^2 V/\partial s_i\partial s_j$.
Also  $\tilde V_{ij}
=\mu^{2\epsilon}\,\big[V_{ij} + 2\epsilon \,\mu^{-2}\,N_{ij}\big]+\cO(\epsilon^2)$,
where 
\medskip
\bea
N_{ij}\equiv \mu\, \Big\{
\,\frac{\partial\mu}{\partial s_i}\, \frac{\partial V}{\partial s_j}
 +\frac{\partial\mu}{\partial s_j}\, \frac{\partial V}{\partial s_i}\Big\}
 +\Big\{\mu\, \frac{\partial^2 \mu}{\partial s_i\partial s_j}
  -\frac{\partial \mu}{\partial s_i}\frac{\partial \mu}{\partial s_j}
\,\Big\}\,V,\qquad   i,j=1,2.
\label{m2prim}
\eea
Then
\bea\label{lt}
V_1 
=\mu(\sigma)^{2\epsilon}
\Big\{
V- \frac{1}{64 \pi^2}\,
\Big[
 \sum_{s=\phi,\sigma} 
M^4_s \,  \Big(\,  \frac{1}{\epsilon}
- \ln\frac{M_s^2(\phi,\sigma)}{c_0\,\mu^2(\sigma)} \Big)
+
\frac{4 \,(V_{ij}\,N_{ji})}{\mu^2(\sigma)}\Big]\Big\}
\eea

\medskip\noindent
with an implicit sum over $i,j$ and with $c_0=4\pi e^{3/2-\gamma_E}$. The one-loop Lagrangian is 
\medskip
\bea\label{l1}
L_1=\frac{1}{2} (\partial_\mu\phi)^2+\frac{1}{2}(\partial_\mu\sigma)^2-V_1.
\eea

\medskip\noindent
Above,  $M_s^2$ denotes
 the field-dependent eigenvalues of the  matrix $V_{ij}$. The poles in $L_1$ are cancelled
by adding the counterterm Lagrangian $\delta L_1$ found using the expression of the  $M_s^2$:
\medskip
\bea\label{ZV}
\delta L_1\equiv -\delta V_1=
- \mu(\sigma)^{2\epsilon} 
\Big\{ \frac{1}{4!} (Z_{\lambda_\phi}\!-1) \lambda_\phi\phi^4+
 \frac{1}{4} (Z_{\lambda_m}\!-1)\lambda_m \phi^2\sigma^2
+\frac{1}{4!} (Z_{\lambda_\sigma}\!-1)\lambda_\sigma\sigma^4\big]
\eea 
with
\bea\label{Z}
Z_{\lambda_\phi}&=&1+\frac{3}{2 \kappa\,\epsilon}  (\lambda_\phi+\lambda_m^2/\lambda_\phi),
\nonumber\\
Z_{\lambda_m}&=&1+\frac{1}{2 \kappa\, \epsilon} (\lambda_\phi+\lambda_\sigma+4\lambda_m),
\nonumber\\
Z_{\lambda_\sigma}&=&1+ \frac{3}{2 \kappa\, \epsilon} (\lambda_\sigma+\lambda_m^2/\lambda_\sigma), 
\quad \kappa=(4\pi)^2.
% \nonumber\\
% Z_{\phi}^{(1)}&=& 1, \quad  Z_{\sigma}^{(1)}=1, \quad \kappa=(4\pi)^2.
\eea

\medskip\noindent
$Z_{\lambda}$'s are identical to their counterparts computed in the same theory 
regularized  with $\mu$=constant (when scale symmetry is broken explicitly). 
The one-loop potential becomes
%\medskip
\bea\label{U}
U_1&= &V+V^{(1)}+V^{(1,n)},
\\[6pt]
V^{(1)} &\equiv & \frac{1}{64\pi^2}\,
\sum_{s=\phi,\sigma} M^4_s(\phi,\sigma)\, \Big[\ln \frac{M^2_s(\phi,\sigma)}{\mu^2(\sigma)}
-\frac{3}{2}\,\Big],
\\
V^{(1,n)}
\!\!\!& \equiv &\!\!\!   \frac{1}{48 \kappa} 
\Big[ \lambda_\phi \lambda_m \frac{\phi^6}{\sigma^2} 
\!-\! (16 \lambda_\phi \lambda_m + 18 \lambda_m^2
\!-\! \lambda_\phi \lambda_\sigma)\phi^4 
\!-\! (48 \lambda_m\! +\! 25 \lambda_\sigma  )\lambda_m\,
 \phi^2 \sigma^2 \!-\! 7 \lambda_\sigma^2 \sigma^4 \Big].
\quad\,\,
\eea

\medskip\noindent
The potential simplifies further if we use the tree-level relation (\ref{minimum})
 among $\lambda_s$ ($s=\phi,m, \sigma$) 
that ensures the spontaneous scale symmetry breaking.
$U_1$ is scale symmetric and a flat direction exists also at the quantum level.
$V_1^{(1,n)}$ is a new, {\it finite} one-loop correction, independent of the parameter $z$;
it contains a {\it non-polynomial} term $\phi^6/\sigma^2$ that can be Taylor-expanded about 
$\langle\phi\rangle$, $\langle\sigma\rangle\not=0$.
$V^{(1,\text{n})}\!\ra\! 0$  in the classical decoupling limit $\lambda_m\!\ra\! 0$. 
The Coleman-Weinberg term is also present, with $\mu\ra \mu(\sigma)$ and thus
 depends on $z$.
This dependence replaces the ``traditional'' dependence of $V^{(1)}$
on the subtraction scale in theories regularized
with $\mu=$constant.   But physics should be independent of this parameter,
which means that $U_1$ must respect the  Callan-Symanzik equation: $dU_1/d\ln z=0$ \cite{tamarit}.

To check the Callan-Symanzik equation,  we need  the beta functions of 
the couplings which run with the momentum, even in scale invariant theories \cite{S3,tamarit}.
 These are computed from the  condition 
$d (\mu(\sigma)^{2\epsilon}\lambda_j Z_{\lambda_j})/d\ln z\!=\!0$ ($j$:\,fixed),
since the bare coupling is independent of $z$. 
The result is identical to that in a theory regularized with $\mu$=constant:
\medskip
\bea
\beta_{\lambda_\phi}^{(1)}&\equiv &\frac{d \lambda_\phi}{d\ln z}
=\frac{3}{\kappa} \,(\lambda_\phi^2+\lambda_m^2),
\\
\beta_{\lambda_m}^{(1)}& \equiv &\frac{d\lambda_m}{d\ln z}=
\frac{1}{\kappa} \,(\lambda_\phi+4\lambda_m+\lambda_\sigma)\lambda_m
\\
\beta_{\lambda_\sigma}^{(1)}& \equiv &\frac{d\lambda_\sigma}{d\ln z}=
\frac{3}{\kappa}\,(\lambda_m^2+\lambda_\sigma^2).
\eea

\medskip\noindent
The Callan Symanzik equation at one-loop is 
\medskip
\bea\label{uui}
\frac{d U_1 (\lambda_j,z)}{d\ln z}=\Big(\beta_{\lambda_j}^{(1)}\,\frac{\partial }{\partial \lambda_j}
+z\,\frac{\partial}{\partial z}\Big)\, U_1(\lambda_j,z)= \cO(\lambda_j^3),
\qquad(\rm{sum\,\,over}\,\, j=\phi,m,\sigma).
\eea

\medskip\noindent
Eq.(\ref{uui})  is easily verified with the above results for the beta functions.
The  one-loop $U_1$ can be used for phenomenology of a  scale invariant version of the
SM extended by the dilaton~\cite{dmg}.

\section{Two-loop potential}

\subsection{ New poles in the two-loop potential}

 To compute the two-loop potential  we use the background field expansion method about 
$\phi,\sigma$. We Taylor-expand $\tilde V$ about these values
\bea
\tilde V(\phi+\delta\phi,\sigma+\delta\sigma)=
V(\phi,\sigma)+ \tilde V_j\, s_j
+\frac12 \tilde V_{jk}\, s_j s_k
+\frac{1}{3!} \tilde V_{ijk}\, s_i s_j s_k
+\frac{1}{4!} \tilde V_{ijkl}\, s_i s_j s_k s_l
+\cdots
\eea

\medskip\noindent
where the subscripts $i, j, k, l$  of $\tilde V_{ij...}$
denote derivatives of $\tilde V$ wrt fields of the set $\{\phi,\sigma\}_j$; 
with $i,j, k,l=1,2$.  Also $s_1=\delta \phi$, $s_2=\delta\sigma$ are field fluctuations. 
Notice
that there are new, evanescent interactions ($\propto\epsilon$)
in vertices $\tilde V_{ijk\cdots}$  generated by eq.(\ref{tildeV})
that impact on the loop corrections.  The two-loop diagrams are presented below. 
Let us first  denote:
%\medskip
 \begin{gather}
 V_2=V_2^a+V_2^b+V_2^c.
 \end{gather}
Then
%\medskip
\bea\label{twoloop}
 V_2^a &=& \frac{i}{12}\; 
\parbox[c][2em][c]{4.0em}{\includegraphics[scale=0.14]{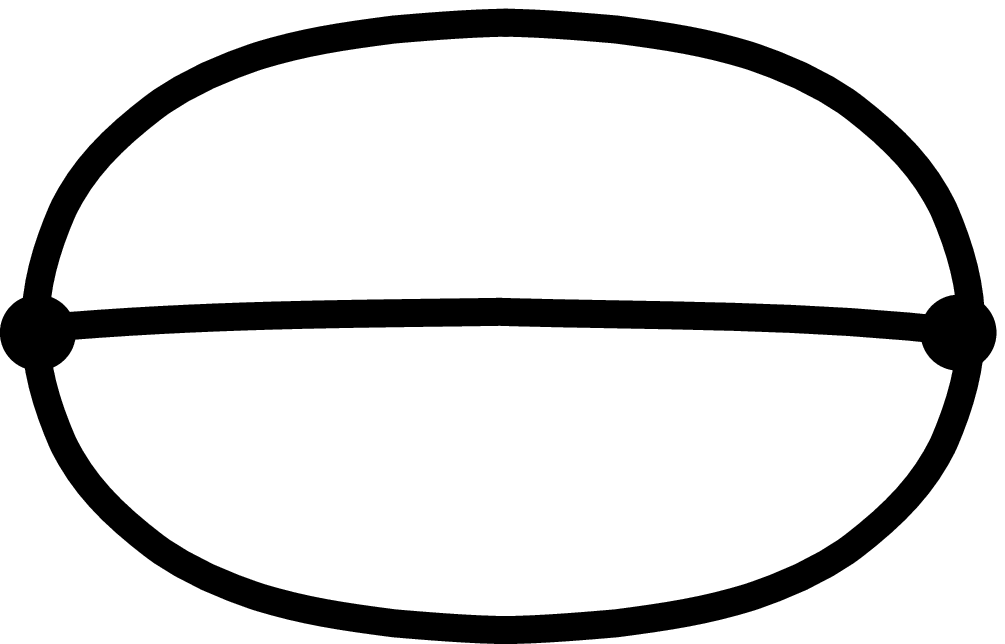}}
=
\frac{i}{12}
\tilde V_{ijk} \tilde V_{lmn}
 \int \frac{d^d p}{(2\pi)^d} \int \frac{d^d q}{(2\pi)^d}
\,(\tilde D_p^{-1})_{il}
\,(\tilde D_q^{-1})_{jm}
\,(\tilde D_{p+q}^{-1})_{kn}
\nonumber\\
&=&
\frac{\mu(\sigma)^{2\epsilon}}{\epsilon^2}\frac{1}{16\kappa^2}
\Big[\phi^4 \,(\lambda_\phi^3+\lambda_\phi\,\lambda_m^2+2\lambda_m^3)
+\sigma^4\,(2\lambda_m^3+\lambda_m^2\lambda_\sigma+\lambda_\sigma^3)
\nonumber\\
&&\hspace{2cm}
+\,\phi^2\,\sigma^2\,(\lambda_\phi^2\,\lambda_m+6 \lambda_\phi\,\lambda_m^2
+10\lambda_m^3+6\lambda_m^2\lambda_\sigma+\lambda_m\lambda_\sigma^2)\Big]
+\cO(1/\epsilon).\qquad
\eea
Also
\bea
V_2^b &=&
 \frac{i}{8}\; 
 \parbox[c][3em][c]{1.9em}{\includegraphics[scale=0.14]{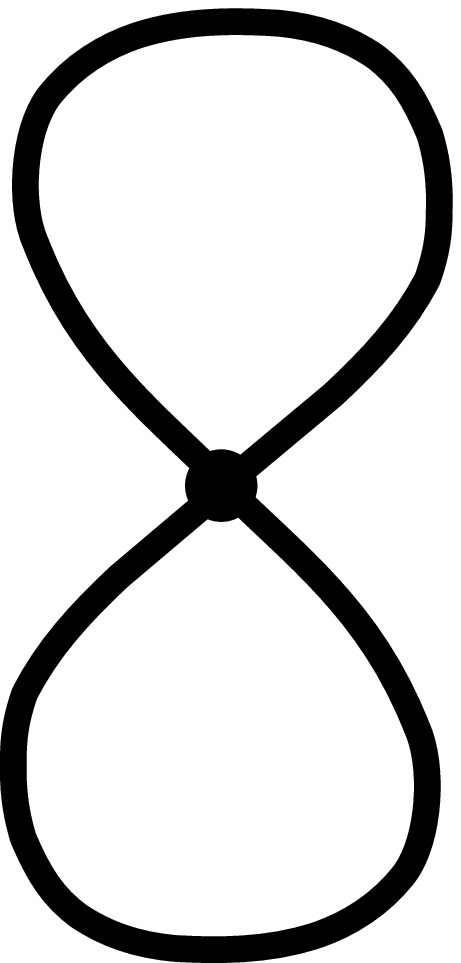}}
=\frac{i}{8}
\tilde V_{ijkl}
\Big[\int \frac{d^d p}{(2\pi)^d} (\tilde D_p^{-1})_{ij}\Big]\,
\Big[ \int \frac{d^d q}{(2\pi)^d}  (\tilde D_q^{-1})_{kl} \Big]
\hspace{3cm}
\nonumber\\
&=& \frac{\mu(\sigma)^{2\epsilon}}{\epsilon^2}\frac{1}{32\kappa^2}
\Big[
\phi^4\,(\lambda_\phi^3+2\lambda_\phi\lambda_m^2+\lambda_m^2\lambda_\sigma)
+\sigma^4\,(\lambda_\phi\lambda_m^2+2\lambda_m^2\lambda_\sigma+\lambda_\sigma^3)
\nonumber\\
& & \hspace{2cm}
+\,2\lambda_m \,\phi^2\sigma^2\,(\lambda_\phi^2+ 9 \lambda_m^2+\lambda_\phi\lambda_\sigma
+\lambda_\sigma^2)\Big]+
\cO(1/\epsilon),\qquad
\eea

\noindent
and finally

\bea\label{dc}
V_2^c & = &  \frac{i}{2}\; 
 \parbox[c][2em][c]{3.2em}{\includegraphics[scale=0.14]{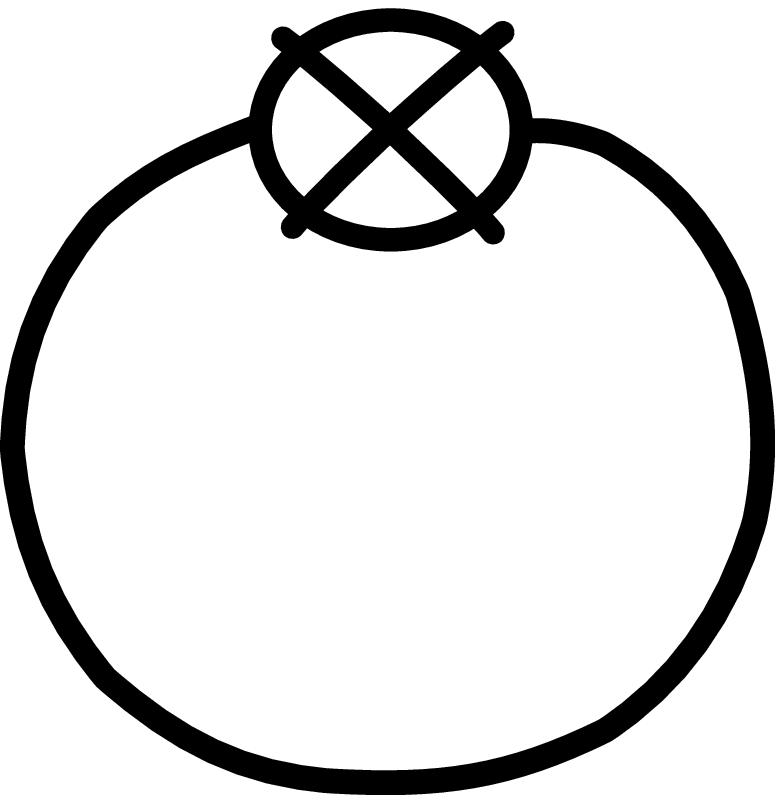}}
=\frac{i}{2}
(\delta V_1)_{ij}\int \frac{d^d p}{(2\pi)^d} \,(\tilde D_p^{-1})_{ij}
\nonumber\\
&=&
\frac{\mu(\sigma)^{2\epsilon}}{\epsilon^2}\frac{(-1)}{16\kappa^2}
\Big[\phi^4\,
(3\lambda_\phi^3+4\lambda_\phi\lambda_m^2+4 \lambda_m^3+\lambda_m^2\lambda_\sigma)
+\sigma^4\,(\lambda_\phi\lambda_m^2+4\lambda_m^3+4\lambda_m^2\lambda_\sigma+3\lambda_\sigma^3)
\nonumber\\
&+&
\phi^2\sigma^2\,(4\lambda_\phi^2\lambda_m+12\lambda_\phi\lambda_m^2+38\lambda_m^3
+2\lambda_\phi\lambda_m\lambda_\sigma+12 \lambda_m^2\lambda_\sigma+4\lambda_m\lambda_\sigma^2)
\Big]+\cO(1/\epsilon).
\eea

\medskip\noindent
These diagrams are computed using \cite{DT}, see also \cite{JJ}.
The propagators are given by  the inverse of the matrix
$(\tilde D_p)_{ij}=p^2\delta_{ij}-\tilde V_{ij}$. To simplify the calculation 
they can be re-written as 
$(\tilde D^{-1})_{ij}=\tilde a_{ij}/(p^2-\tilde V_p)+\tilde b_{ij}/(p^2-\tilde V_m)$,
with appropriate coefficients $\tilde a_{ij}$ and $\tilde b_{ij}$ and
where $\tilde V_p$, $\tilde V_m$ ($\tilde V_p>\tilde V_m$)
denote the field-dependent masses, eigenvalues of the matrix 
$\tilde V_{ij}$, $i,j=\phi,\sigma$. Note that $\tilde V_{ij}$, $\tilde V_p$, $\tilde V_m$,
$\tilde a_{ij}$, $\tilde b_{ij}$ and also $\tilde V_{ijkl}$, $\tilde V_{ijk}$
 contain positive powers of $\epsilon$; 
this is relevant for the above calculation, since they contribute 
to the finite and $1/\epsilon$ parts of the potential.
Their form is detailed in Appendix~\ref{appendixB} and \ref{appendixC}.

One  notices that the  poles $1/\epsilon^2$ in $V_2^{a,b,c}$ are identical to those in the
theory regularized with $\mu=$constant. This is expected for this
 leading singularity, but this is not true for their sub-leading one ($1/\epsilon$) or
 for their finite part (see later).
The long expressions $\cO(1/\epsilon)$ and $\cO(\epsilon^0)$ of each diagram $V_2^{a,b,c}$
are not shown here.   The sum of these diagrams gives
\medskip
\bea\label{V2}
\!\!\!\!V_2&=& 
\frac{\mu(\sigma)^{2\epsilon}}{\epsilon^2}\frac{(-1)}{32 \kappa^2} 
\,\Big[\phi^4 (3 \lambda_\phi^3 
+\! 4 \lambda_\phi \lambda_m^2\! +  4\lambda_m^3 + \lambda_m^2 \lambda_\sigma)
+\!\sigma^4(3 \lambda_\sigma^3\! + {\lambda_\phi \lambda_m^2}
 +\! 4 \lambda_m^3+ 4\lambda_m^2 \lambda_\sigma) 
\nonumber \\
&& \hspace{1.7cm} +\,\phi^2 \sigma^2 (4 \lambda_\phi^2 \lambda_m
 + 12 \lambda_\phi \lambda_m^2 + 38 \lambda_m^3+
 2 \lambda_\phi \lambda_m \lambda_\sigma + 12\lambda_m^2 \lambda_\sigma+ 4\lambda_m \lambda_\sigma^2)
\Big]
\nonumber\\
&+& \frac{\mu(\sigma)^{2\epsilon}}{\epsilon}\; \frac{1}{16 \kappa^2} 
\,\Big[\phi^4 ( \lambda_\phi^3 
+  \lambda_\phi \lambda_m^2 +  2\lambda_m^3 ) +\sigma^4(2 \lambda_m^3+ \lambda_m^2 \lambda_\sigma+\lambda_\sigma^3)
\\
&&\hspace{1.7cm} +\phi^2 \sigma^2 (\lambda_\phi^2 \lambda_m 
+ 6 \lambda_\phi \lambda_m^2 + 10 \lambda_m^3 + 6 \lambda_m^2 \lambda_\sigma+ \lambda_m \lambda_\sigma^2) 
\Big] +V_2^{1/\epsilon}\!\!+V^{(2)}+V^{(2,n)}.
\nonumber
\eea

\medskip\noindent
Here $V^{(2)}$ and $V^{(2,n)}$ are $\cO(\epsilon^0)$ i.e. finite  quantum 
corrections presented in Appendix~\ref{appendixB}.
 $V_2^{1/\epsilon}=\cO(1/\epsilon)$ is a {\it new term} that contains $1/\epsilon$ poles
{\it not} present in the theory regularized with $\mu$=constant;
its origin is  due to evanescent interactions ($\propto\epsilon$), which ``meet''
$1/\epsilon^2$ poles, thus giving $1/\epsilon$ terms.
One finds
\medskip
\bea\label{opp}
\!\!\! V_2^{1/\epsilon}
\!\!&=&
\frac{\mu(\sigma)^{2\epsilon}}{16 \kappa^2\,\epsilon} \,
\Big[
\phi^4 \Big(\frac{20}{3}\lambda_\phi^2 \lambda_m \!+  \frac{7}{6} 
\lambda_\phi \lambda_m^2\! - 2\lambda_m^3\! - \frac{1}{2} \lambda_\phi^2 \lambda_\sigma
\! - \frac{4}{3}\lambda_\phi \lambda_m \lambda_\sigma 
\!+ \frac{7}{12} \lambda_m^2 \lambda_\sigma\! +\!
 \frac{1}{4} \lambda_\phi \lambda_\sigma^2
 \Big) 
\nonumber\\
&+&\!\!
\phi^2 \sigma^2 \Big(\,8 \lambda_\phi \lambda_m^2 
+\frac{41}{2} \lambda_m^3 + \lambda_\phi \lambda_m \lambda_\sigma 
+ \frac{43}{3}\lambda_m^2\lambda_\sigma +\frac{1}{2} \lambda_m \lambda_\sigma^2 \Big) 
% \nonumber\\
+ 
\sigma^4\Big(4 \lambda_m^3\!+ \frac{1}{3}\lambda_m^2 \lambda_\sigma
\!+\frac{7}{4}\lambda_\sigma^3 \Big) 
\nonumber\\
& +&
 \frac{\phi^6}{\sigma^2}\, \Big( -\frac{7}{6} \lambda_\phi^2 \lambda_m 
+ \frac{7}{3} \lambda_\phi \lambda_m^2  - \frac{1}{6} \lambda_\phi \lambda_m \lambda_\sigma \Big) 
-
 \frac{1}{4} \lambda_\phi \lambda_m^2 \,\,\frac{\phi^8}{\sigma^4}\, \Big]. 
\eea

\bigskip\noindent
In addition to usual counterterms ($\phi^4$, etc), 
notice from eq.(\ref{opp}) the need for  {\it non-polynomial} 
counterterms $\phi^6/\sigma^2$ and $\phi^8/\sigma^4$
(see  also $\phi^6/\sigma^2$ in eq.(\ref{U})).
The above two-loop results contribute to the Lagrangian 
(below $\rho^{\phi},\rho^\sigma$ are  wavefunction coefficients defined later)
%\medskip
\bea\label{l2}
L_2=\frac{1}{2}\Big( \frac{\rho^\phi}{\epsilon}+\text{finite}\Big) (\partial_\mu\phi)^2
+
\frac{1}{2}\Big(\frac{\rho^\sigma}{\epsilon}+\text{finite}\Big) (\partial_\mu\sigma)^2
-V_2.
\eea

\medskip\noindent
A counterterm $\delta L_2$ cancels the poles in  the sum $L_1+L_2$ of
eqs.(\ref{l1}), (\ref{l2}) up to two-loops
\medskip
\bea\label{deltaL2}
\delta L_2&=&
\frac{1}{2}(Z_\phi-1)(\partial_\mu\phi)^2
+
\frac{1}{2}(Z_\sigma-1)(\partial_\mu\sigma)^2
-
 \mu(\sigma)^{2\epsilon}\,
\Big\{\, (Z_{\lambda_\phi}-1)\frac{\lambda_\phi}{4!}\phi^4 +
\nonumber\\
&+&
\!\!\!\! (Z_{\lambda_m}-1)\frac{\lambda_m}{4}\phi^2\,\sigma^2
\,+\,(Z_{\lambda_\sigma}-1)\frac{\lambda_\sigma}{4!} \sigma^4
\!+\! (Z_{\lambda_6}-1)\frac{\lambda_6}{6} \frac{\phi^6}{\sigma^2}
\!+\! (Z_{\lambda_8}-1)\frac{\lambda_8}{8} \frac{\phi^8}{\sigma^4}
\,\Big\},\,\,\,
\eea 
%\medskip\noindent
where 
\bea\label{Z1}
Z_{\lambda_\phi}&=&1+\frac{\delta_0^\phi}{\kappa\,\epsilon}+\frac{1}{\kappa^2} 
\Big(\,\frac{\delta_1^\phi+\nu_1^\phi}{\epsilon}+\frac{\delta_2^\phi}{\epsilon^2}\,\Big),
\nonumber\\
Z_{\lambda_m}&=&1+\frac{\delta_0^m}{\kappa\,\epsilon}+
\frac{1}{\kappa^2}\Big(\,\frac{\delta_1^m+\nu_1^m}{\epsilon}+\frac{\delta_2^m}{\epsilon^2}\,\Big),
\nonumber\\
Z_{\lambda_\sigma}&=&1+ 
\frac{\delta_0^\sigma}{\kappa\,\epsilon}+
\frac{1}{\kappa^2}\Big(\,\frac{\delta_1^\sigma+\nu_1^\sigma}{\epsilon}
+\frac{\delta_2^\sigma}{\epsilon^2}\,\Big),
\nonumber\\
Z_{\lambda_6} &=& 1 + \frac{1}{\kappa^2} \frac{\nu_1^6}{\epsilon};
\qquad
Z_{\lambda_8}= 1 + \frac{1}{\kappa^2} \frac{\nu_1^8}{\epsilon},
\eea

\medskip\noindent
where one-loop  $\delta_0^\phi$, $\delta_0^m$, $\delta_0^\sigma$ can be read from eq.(\ref{Z})
while the two-loop coefficients  $\delta_{k}^{s}$, $k=1,2$, $s=\phi, m, \sigma$,
are shown in  Appendix~\ref{appendixA}. They are obtained by comparing 
$\delta L_2$ against $L_2$, using $V_2$ of eq.(\ref{V2}).
 The  coefficients  $\delta_k^s$ are those of the theory regularized
with $\mu=$constant. However, there is  an extra  contribution from coefficients
 $\nu_1^s$,  $s=\phi,m,\sigma, 6,8$ (see Appendix~\ref{appendixA}),
  that is generated by the new poles $1/\epsilon$ of
$V_2^{1/\epsilon}$. This new contribution  brings a correction to the two-loop
 beta functions of our theory, see later.

One can also  show that the two-loop-corrected wavefunction coefficients have 
expressions similar to those in the theory regularised with $\mu=$constant:
\medskip
\bea
\label{Z2}
 Z_\phi &=& 1 +\frac{\rho^\phi}{\kappa^2\epsilon},
\qquad\quad
\rho^\phi= - \frac{1}{24}\, (\lambda_\phi^2 + 3 \lambda_m^2),
\nonumber\\
Z_\sigma &=& 1 +\frac{\rho^\sigma}{\kappa^2\,\epsilon},
\qquad\quad
\rho^\sigma=  - \frac{1}{24}\, (\lambda_\sigma^2 + 3 \lambda_m^2),
\eea 
One often uses the notation $\gamma_\phi=-2\rho^\phi/\kappa^2$ and
 $\gamma_\sigma=-2\rho^\sigma/\kappa^2$  for the anomalous dimensions.

\subsection{Two-loop beta functions}

With the above information, one obtains the two-loop beta functions. To this purpose, one
uses that the  ``bare'' couplings $\lambda^B_j$ below are independent of the parameter $z$:
%\medskip
\bea\label{b}
\lambda_\phi^B &=& \mu(\sigma)^{2\epsilon}\lambda_\phi\,Z_{\lambda_\phi}\,Z_\phi^{-2},
\nonumber\\
\lambda_m^B &=& \mu(\sigma)^{2\epsilon}\lambda_m\,Z_{\lambda_m}\,Z_\phi^{-1}\,Z_\sigma^{-1},
\nonumber\\
\lambda_\sigma^B &=& \mu(\sigma)^{2\epsilon}\lambda_\sigma\,Z_{\lambda_\sigma}\,Z_\sigma^{-2},
\nonumber\\
\lambda_6^B &=& \mu(\sigma)^{2\epsilon}\lambda_6\,Z_{\lambda_6}\,Z_\phi\,Z_\sigma^{-3}.
\eea

\medskip\noindent
We thus demand that   $(d/d\ln z) \lambda_k^B=0$, $k=\phi,m,\sigma, 6, 8$ \footnote{
We also include the effect of
 wavefunction renormalization  of the subtraction function which demands replacing:
$\mu(\sigma)=z\,\sigma^{1/(1-\epsilon)}\ra z\,(Z_\sigma^{1/2}\,\sigma)^{1/(1-\epsilon)}$; however, 
 this brings no correction in this order.}.
Taking the logarithm  of the first expression in (\ref{b})
and then the derivative  with respect to $\ln z$, one obtains 
\medskip
\bea
2\epsilon+\frac{\beta_{\lambda_\phi}}{\lambda_\phi}
+\sum_{j=\phi,m,\sigma}\,\beta_{\lambda_j}\,\frac{d}{d\ln z}
 \ln \Big[Z_{\lambda_\phi} Z_\phi^{-2}\Big]=0
\eea
and similar expressions for the other couplings. Using the form of $Z'$s, one finds
\medskip
\bea\label{bb}
\beta_{\lambda_\phi}
=
-2\epsilon\lambda_\phi
+
2\lambda_\phi \sum_{j=\phi,m,\sigma}
\lambda_j\,\frac{d}{d\lambda_j}
\Big(\frac{\delta_0^\phi}{\kappa}+\frac{\delta_1^\phi+\nu_1^\phi}{\kappa^2}
-\frac{2\,\rho^\phi}{\kappa^2}\Big).
\eea

\medskip\noindent
One easily obtains similar relations  for $\beta_{\lambda_m}$ and $\beta_{\lambda_\sigma}$ 
(for $\beta_{\lambda_\sigma}$ just replace the sub-/super-script $\phi\ra \sigma$).
The difference in these beta functions 
from those in the  same  theory but regularized with $\mu=$constant is the presence 
 of a new contribution:  $\nu_1^{\phi}$ ($\nu_1^m$, $\nu_1^\sigma$, respectively), 
 that we identified in eqs.(\ref{Z1}).
Eq.(\ref{bb}) is solved with particular attention to the 
$\epsilon$-dependent terms, to find at two-loop:
\medskip
\bea
\beta_{\lambda_\phi}&=&\frac{3}{\kappa}(\lambda_\phi^2+\lambda_m^2)
-\frac{1}{\kappa^2}(\frac{17}{3}\laf^3 + 5\laf\lam^2 + 12 \lam^3)
+\beta_{\laf}^{(2,\text{n})},
\nonumber\\
\beta_{\lam} &=&\frac{1}{\kappa}(\laf + 4\lam + \las)\lam
 - \frac{\lam}{6\kappa^2} (5\laf^2 + 36 \laf \lam + 54 \lam^2 + 36 \lam \las + 5\las^2)
+ \beta_{\lam}^{(2,\text{n})},
\nonumber\\
\beta_{\las} &=& \frac{3}{\kappa}(\lambda_m^2+\lambda_\sigma^2)-
\frac{1}{\kappa^2}(12\lam^3 + 5\lam^2 \las + \frac{17}{3}\las^3)
+\beta_{\las}^{(2,\text{n})}.
\eea

\medskip\noindent
The  ``new'' terms  $\beta_\lambda^{(2,n)}$  on the rhs are
\bea\label{betas2}
 \beta_{\laf}^{(2,\text{n})} &=& \frac{1}{2 \kappa^2}
\left[ \lam^2 (24\lam - 7 \las)+ \laf(-14 \lam^2 + 16 \lam \las -3 \las^2) 
+ \laf^2(-80\lam + 6 \las) \right],
\nonumber\\
\beta_{\lam}^{(2,\text{n})} &=& -\frac{\lam}{6 \kappa^2} \,(48 \laf \lam + 6 \laf \las + 
123 \lam^2 + 86 \lam \las + 3 \las^2),
\nonumber\\
\beta_{\las}^{(2,\text{n})} &=& -\frac{1}{2 \kappa^2}\,(48\lam^3 + 4 \lam^2 \las + 21 \las^3),
\nonumber\\
\beta_{\lambda_6}^{(2,\text{n})} &=& \frac{1}{4\kappa^2}\, \laf \lam (7\laf - 14 \lam + \las),
\nonumber\\
\beta_{\lambda_8}^{(2,\text{n})} &=& \frac{1}{2\kappa^2}\, \laf \lam^2.
\eea

\medskip\noindent
Here  $\beta^{(2,\text{n})}_\lambda$  that appears for each $\lambda$ at two-loop is the 
mentioned correction, that is missed if this theory is regularized 
with $\mu=$constant, when one  breaks explicitly the scale symmetry. 
Notice that $\lambda_{6,8}$ also run in this order in the scale invariant theory.

We conclude that  from the two-loop running of 
the couplings, encoded by the beta functions,
 one can distinguish between the theory with (spontaneously 
broken) scale symmetry at quantum level and that in  which this symmetry is broken 
{\it explicitly}  by quantum corrections (with $\mu=$constant). 
There is a simple way to understand this difference:
the theory regularized with $\mu=$constant, and two fields $\phi$, $\sigma$ is renormalizable
while our model, scale invariant at quantum level, is non-renormalizable.
This is due to the scale-invariant  non-polynomial terms of type $\phi^6/\sigma^2$,
 $\phi^8/\sigma^4$ generated at one- and  two-loop level\footnote{
 This non-renormalizability argument is different
from that in \cite{S2} which does not apply here, see \cite{dmg}.}.
This justifies the different beta functions in the two approaches starting from
the two-loop level.
This is an interesting result of the paper.

\subsection{Two-loop potential after renormalization}

Finally, we  present the two-loop potential $U$ after renormalization. It has the form
\medskip
\bea\label{finalU}
U=\underbrace{V+V^{(1)}+V^{(1,\text{n})}}_{=U_1}+V^{(2)}+V^{(2,\text{n})}
\eea

\medskip\noindent
where $U_1$ is the one-loop result of  (\ref{U}).
$V^{(2)}$ is a two-loop correction identical   to that obtained in the theory regularized with
$\mu=$constant (up to  replacing $\mu\ra z\, \sigma)$,
while $V^{(2,\text{n})}$ are new two-loop terms that involve derivatives of $\mu(\sigma)$ wrt 
$\sigma$ (similar to  one-loop $V^{(1,\text{n})}$)\footnote{See \cite{EJ} for 
 further discussion on the Goldstone modes contributions to the potential.}
The long expressions of $V^{(2)}$,  $V^{(2,n)}$ are given
in Appendix~\ref{appendixB}, eq.(\ref{B5}).
$U$ contains  new, {\it non-polynomial} effective operators,  
such as $\phi^6/\sigma^2$ and $\phi^8/\sigma^4$, etc:
%\medskip
\bea\label{uu}
U=
\frac{7 \lambda_\phi^3}{576 \, \kappa^2}\frac{\phi^8}{\sigma^4}+
\frac{5\lambda_\phi^3}{24 \,\kappa^2}\frac{\phi^6}{\sigma^2}
+\cdots
\eea

\medskip\noindent
All non-polynomial terms present in the potential can be expanded about the ground state 
\bea
\phi=\langle\phi\rangle+\delta\phi,\qquad  
\sigma=\langle\sigma\rangle+\delta\sigma
\eea
where
$\delta\phi$ and $\delta\sigma$ represent fluctuations about the ground state. 
Then each non-polynomial operator becomes an infinite series expansion about the
point  $\langle\phi\rangle/\langle\sigma\rangle$. For example
\medskip
\bea\label{exp}
\frac{\phi^6}{\sigma^2}=(\langle\phi\rangle+\delta\phi)^4\,\,
 \frac{\langle\phi\rangle^2}{\langle\sigma\rangle^2}\,\,
\Big(1+\frac{2\delta \phi}{\langle\phi\rangle}+\frac{\delta\phi^2}{\langle\phi\rangle^2}+\cdots\Big)
\,\,
\Big(1-\frac{2\delta \sigma}{\langle\sigma\rangle}+\frac{3\delta\sigma^2}{\langle\sigma\rangle^2}+\cdots\Big)
\eea

\medskip\noindent
and similarly for the operator $\phi^8/\sigma^4$ in $U$, etc.
Although we did not present  the ground state of the one-loop potential, this
 is known to satisfy the relation \cite{dmg}:
$\langle\phi\rangle^2/\langle\sigma\rangle^2=
-3\lambda_m/\lambda_\sigma (1+\text{loop-corrections})$ \cite{dmg}.
Using this information in eq.(\ref{exp}) and (\ref{uu}), one sees that in the classical
decoupling limit $\lambda_m\ra 0$, the non-polynomial operators of (\ref{uu}) do vanish.

It is important to stress that only operators 
of the form $\phi^{2n+4}/\sigma^{2n}$, $n\geq 1$ were generated in the two-loop potential, 
but no operator  like  $\sigma^{2n+4}/\phi^{2n}$, $n\geq 1$ is present.
This is due to the way the subtraction function enters in the loop corrections,
via derivatives wrt $\sigma$ of  $\mu(\sigma)^\epsilon$ which are suppressed by 
positive powers of $\mu(\sigma)$.  This means that all higher dimensional operators
are ultimately suppressed  by (large) $\langle\sigma\rangle$ and not proportional to it.
This is welcome for the  hierarchy problem, since such terms could otherwise lead
to corrections to the higgs mass of the type  $\lambda_\phi^3 \langle\sigma\rangle^2$ 
requiring tuning the higgs self-coupling $\lambda_\phi$, and thus re-introducing the hierarchy problem.
This problem is avoided at least at one-loop \cite{S1,dmg}.

\section{Two-loop Callan-Symanzik for the potential}

A good check of our  two-loop scale-invariant potential is 
the Callan-Symanzik equation, in its version for scale invariant theories
\cite{tamarit}. This equation states the independence of the two-loop potential of
the  subtraction (dimensionless) parameter $z$; this parameter fixes the subtraction scale to
$z\langle\sigma\rangle$, after spontaneous scale symmetry breaking. 
The equation is 
\medskip
\bea \label{CS}
\frac{d\, U(\lambda,z)}{d \ln z}=
\Big(
z\frac{\partial}{\partial z}
+
\beta_{\lambda_j}\, \frac{\partial}{\partial \lambda_j} 
-
\phi\gamma_\phi\,\frac{\partial}{\partial\phi}
-
\sigma\gamma_\sigma\,\frac{\partial}{\partial\sigma}\Big) U(\lambda,z)
=0\;,
\eea
where the $j$-summation runs over $\lambda_j=\laf,\lam,\las,\lambda_6,\lambda_8$.
Eq.\eqref{CS} can be re-written as a 
set of equations at a given order of $\lambda$'s (or number of loops).
To help one trace the difference between our scale-invariant result  and that 
of the same theory but with  $\mu=$constant, below we use for $U$ the 
decomposition given in eq.(\ref{finalU}) while  for the beta functions
we write
\medskip
\bea
\beta_{\lambda_j} &=& \beta^{(1)}_{\lambda_j} + \beta^{(2)}_{\lambda_j}+ \beta^{(2,\text{n})}_{\lambda_j}.
\eea

\medskip\noindent
The terms in the beta function
correspond to 1-loop ($\beta_\lambda^{(1)}$), 2-loop-only ($\beta_\lambda^{(2)}$)
and 2-loop-new parts ($\beta_\lambda^{(2,\text{n})}$).
Then, with the two-loop anomalous dimensions $\gamma_\phi$, $\gamma_\sigma$ defined 
after eq.(\ref{Z2}),  a careful analysis shows that
 eq.\eqref{CS}  splits into 
\bea
&& \frac{\partial \, V^{(1)}}{\partial \ln z} \,
 + \,  \beta^{(1)}_{\lambda_j}\, \frac{\partial V}{\partial \lambda_j}  =0 \label{CSOLD1}
\\
&& \frac{\partial \, V^{(1,\text{n})}}{\partial \ln z}  = 0 \label{CSNEW1} 
\\
&& \frac{\partial \, V^{(2)} }{\partial \ln z} 
 + \left( \beta^{(2)}_{\lambda_j} 
\frac{\partial \;}{\partial \lambda_j} - \gamma_\phi \phi 
\frac{\partial}{\partial \phi}- \gamma_\sigma \sigma \frac{\partial}{\partial \sigma}\right) 
V +  \beta^{(1)}_{\lambda_j} \,\frac{\partial V^{(1)}}{\partial\lambda_j}=0 \label{CSOLD2}
\\
&&
\frac{\partial\, V^{(2,\text{n})}}{\partial \ln z} +  \beta^{(2,\text{n})}_{\lambda_j} 
\frac{\partial  V}{\partial \lambda_j} +  \beta^{(1)}_{\lambda_j}\,
\frac{\partial V^{(1,\text{n})}}{\partial \lambda_j}=0 \label{CSNEW2} \;.
\eea

\medskip\noindent
where $V$  includes the new terms
$(\lambda_6/6)\, \phi^6/\sigma^2+ (\lambda_8/8)\, \phi^8/\sigma^4$.
We checked that these equations are respected. 
Eqs.\eqref{CSOLD1}, \eqref{CSOLD2} express the \textit{usual} Callan-Symanzik equation (of the theory
with $\mu=$constant), whereas \eqref{CSNEW1} and \eqref{CSNEW2} constitute a \textit{new} part,
 which is nonzero only when $\mu=\mu(\sigma)$.
Eq.\eqref{CSNEW1} is obvious and hardly revealing. But 
checking eq.\eqref{CSNEW2} is more difficult. 
For this one also needs to take account of the ``new'' corrections to
two-loop beta functions of $\lambda_{6,8}$, see eq.(\ref{betas2}) and also 
the  $z$-dependent part of $V^{(2,\text{n})}$ which we write below
\smallskip
\bea
V^{(2,\text{n})}& =& \frac{1}{192\,\kappa^2}\Big[ 
\overline{\ln}V_p + \overline{\ln}V_m\Big]\Big[
\big( -144 \laf^2 \lam 
- 111 \laf \lam^2 - 168 \lam^3 + 9 \laf^2 \las - 40 \lam^2 \las \big)\, \phi^4 
\nonumber\\
&-& \big(192 \laf \lam^2 + 705 \lam^3 + 
 37 \laf \lam \las + 368 \lam^2 \las + 106 \lam \las^2\big)\,\phi^2 \sigma^2 
\nonumber\\
 & -& \big(48 \lam^3 + 46 \lam^2 \las + 63 \las^3\big)\,\sigma^4 + 
\big(  18 \laf^2 \lam - 24 \laf \lam^2 + 3 \lam^3 + 
 3 \laf \lam \las \big)\, \frac{\phi^6}{\sigma^2} 
\nonumber\\
& + & 3 \laf \lam^2\, \frac{\phi^8}{\sigma^4}
 \Big] + \text{$z$-independent terms}, \quad
\text{where}\quad
\overline\ln A\equiv\ln\frac{A}{(z\sigma)^2 4\pi e^{-\gamma_E}}-1.
\eea

\smallskip\noindent
where $\gamma_E=0.5772....$.
Here $V_p$ and $V_m$ are field dependent eigenvalues of the matrix of second derivatives 
$V_{ij}$ wrt $i,j=\phi,\sigma$ of the tree level potential. Given this, the
Callan-Symanzik equation of the potential is verified at the two-loop level.

\section{Conclusions}

Quantum scale invariance with spontaneous breaking may provide 
a solution to the cosmological constant 
and the hierarchy problem. The ``traditional'' method for loop calculations
 breaks explicitly classical scale symmetry of a theory
 due to the regularization which introduces a  subtraction scale (DR scale, cut-off, 
Pauli-Villars scale). However, it is known how
to perform quantum calculations in a manifestly scale invariant way: the
 subtraction scale is replaced by a subtraction  function of the field(s) (dilaton $\sigma$)
 which when acquiring a vev spontaneously,  generates this scale
 $\mu(\langle\sigma\rangle)=z\langle\sigma\rangle$. The Goldstone mode of this  
symmetry is the dilaton field which remains a flat direction of the 
quantum scale-invariant potential.

Starting with a classically scale-invariant action, 
we computed the two-loop  scalar potential
of $\phi$ (higgs-like) and $\sigma$  in a scale invariant regularization.
The one- and two-loop potential are scale invariant and contain new terms
beyond the usual corrections obtained for $\mu=$constant (Coleman-Weinberg, etc),
due to field derivatives of $\mu(\sigma)$. They also contain
interesting  effective {\it non-polynomial} 
 operators $\phi^6/\sigma^2$ and $\phi^8/\sigma^4$, etc, allowed by scale symmetry,
showing that such theories are non-renormalizable.
 These operators can be expanded about the non-zero  $\langle\phi\rangle$
 and $\langle\sigma\rangle$, 
to obtain an infinite series of  effective polynomial ones, 
suppressed by $\langle\sigma\rangle\gg\langle\phi\rangle$
(such hierarchy can be enforced by one  initial, classical tuning of  the couplings). 
 The non-polynomial operators emerge
from evanescent interactions  ($\propto\epsilon$)  between $\phi$ 
and $\sigma$ that vanish in $d=4$ but
are demanded by scale invariance in $d=4-2\epsilon$.
Previous works also showed that the  higgs mass is stable against  quantum corrections 
 at one-loop $m_\phi^2\ll \langle\sigma\rangle^2$,
which may remain true beyond it  if only spontaneous scale symmetry breaking  is present.

 We checked the consistency of the two-loop scale invariant potential 
by showing that it satisfies  the Callan-Symanzik equation in its scale-invariant formulation.
To this purpose we computed the two-loop beta functions of the couplings of the theory.
While one-loop beta functions are exactly  those of  the same theory of $\phi$, $\sigma$
regularized with $\mu\!=$constant,  the two-loop beta functions
 differ from those of the theory regularized  with explicit breaking of scale symmetry 
($\mu=$constant). In conclusion, the running of the couplings enables one to distinguish 
between spontaneous and explicit breaking of quantum scale  symmetry of the action.

\begin{center}
--------------------------------
\end{center}

\section*{Appendix:}

\def\theequation{A-\arabic{equation}}
\def\thesubsection{A}
\setcounter{equation}{0}

 \subsection{Coefficients of the counterterms}\label{appendixA}
 
Assuming $\lambda_n=0$, $n \geqslant 6$ at tree-level,
 the coefficients of eqs.(\ref{Z1}), (\ref{bb}) are:
 \bea
\delta_1^\phi &=& - \frac{3}{2} \left( \laf^2 + \lam^2 + 2 \frac{\lam^3}{\laf}\right) \\
\delta_1^m &=& - \frac{1}{4} \left( \laf^2 + 6 \laf\lam + 10\lam^2 + 6\lam \las + \las^2 \right) \\
\delta_1^\sigma &=& -\frac{3}{2} \left( 2\frac{\lam^2}{\las} + \lam^2 + \las^2 \right) \\
\nu_1^\phi &=& \frac{1}{8} \left[ \frac{\lam^2}{\laf}(24 \lam - 7 \las)
 -(14\lam^2 - 16 \lam \las + 3 \las^2)-\laf(80\lam - 6 \las) \right] \\
\nu_1^m &=& - \frac{1}{24} \left[123 \lam^2 + 86 \lam \las +3 \las^2 + 6\laf(8 \lam + \las)\right] \\
\nu_1^\sigma &=& -\frac{1}{8 \lambda_\sigma} \left( 48 \lam^3 + 4 \lam^2 \las + 21 \las^3 \right) \\
\nu_1^6 &=& \frac{1}{16} \frac{\laf \lam}{\lambda_6} \left(7 \laf - 14 \lam + \las \right) \\
\nu_1^8 &=& \frac{1}{8} \frac{\laf \lam^2}{\lambda_8} \\
\delta_2^\phi &=& \frac{3}{4} \left[ 3\laf^2 + 4\lam^2 + \frac{\lam^2}{\laf} (4 \lam + \las)\right] \\
\delta_2^m &=& \frac{1}{4}\left(2\laf^2 + 6\laf \lam + \laf \las + 19 \lam^2 + 6\lam \las + 2 \las^2 \right) \\
\delta_2^\sigma &=& \frac{3}{4}\left[ \frac{\lam^2}{\las} (\laf + 4 \lam) + 4\lam^2 + 3 \las^2 \right].
\eea

%\newpage
 \def\theequation{B-\arabic{equation}}\def\thesubsection{B}
\subsection{The finite part of the two-loop potential}
\setcounter{equation}{0}
 \label{appendixB}

We provide here the finite part of the two-loop potential, 
$V^{(2)}+V^{(2,n)}$ of eq.(\ref{V2}), (\ref{finalU}).
This is rather long,  we thus use a simplified notation.
The propagators  are found  from:
$(\widetilde D_p)_{ij}=p^2\delta_{ij}-\widetilde V_{ij}$. 
To simplify the calculation it helps to write them as
%\medskip
\bea
&&(\widetilde D^{-1})_{ij} = \frac{\tilde a_{ij}}{p^2-\widetilde V_p}
+\frac{\tilde b_{ij}}{p^2-\widetilde V_m}, 
\qquad 
\tilde b_{ij}=\delta_{ij}-\tilde a_{ij}
\nonumber\\
\tilde a_{11}\!\! &=&\!\!
 \tilde b_{22} = \frac{\widetilde V_p-\widetilde V_{22}}{\widetilde V_p-\widetilde V_m},\quad 
\tilde a_{22} = \tilde b_{11} = 1 - \tilde a_{11} 
=\frac{\widetilde V_p-\widetilde V_{11}}{\widetilde V_p-\widetilde V_m},\quad 
\tilde a_{12}=\tilde a_{21}=\frac{\widetilde V_{12}}{\widetilde V_p -\widetilde V_m},
\quad
\eea

\smallskip\noindent
where $\widetilde V_p$, $\widetilde V_m$ are the field-dependent eigenvalues of matrix 
$\widetilde V_{ij}=\partial^2
 \widetilde V/\partial s_i\partial s_j$, $i,j=1,2; \,\,s_1=\phi, s_2=\sigma$,
and $\widetilde V=\mu(\sigma)^{2\epsilon}\,V$ where $V$ is the tree level potential in $d=4$.
We introduce the following coefficients (without \, $\widetilde{}$\,\,)
of the Taylor expansions in $\epsilon$  
(see Appendix~\ref{appendixC} for their values in terms of the couplings and fields):
\bea\label{intro}
\tilde a_{ij}&=&a_{ij}+\epsilon\,a_{ij}^1+\epsilon^2\,a_{ij}^2+\cO(\epsilon^3),
\qquad
b_{ij}=\delta_{ij}-a_{ij}, \quad b_{ij}^1=-a_{ij}^1, \quad b_{ij}^2=-a_{ij}^2
\nonumber\\
\widetilde V_{ijk\ldots}\!\!\!\!&=&\!\!\!
\mu(\sigma)^{2\epsilon}\, \big[\,
v_{ijk\ldots}+\epsilon\,u_{ijk\ldots}+\epsilon^2\,w_{ijk\ldots}+\cO(\epsilon^3)\,\big],
\qquad \text{where:}
\nonumber\\
\widetilde V_{ijk\ldots}\!\!&=&\!\!
\frac{\partial^4 \widetilde V}{\partial s_i\partial s_j \partial s_k \cdots},
\quad
v_{ijk\ldots}=
\frac{\partial^4 V}{\partial s_i \partial s_j \partial s_k \cdots},
\quad
i,j,k,\ldots=1,2;
\quad
s_1\!=\!\phi,\, s_2\!=\!\sigma,
\nonumber\\
\widetilde V_p&=&\mu(\sigma)^{2\epsilon}\, V_p \,\big[
1+ c_p^1\,\epsilon +c_p^2\,\epsilon^2+\cO(\epsilon^3)\big]
\nonumber\\ 
\widetilde V_m&=&\mu(\sigma)^{2\epsilon}\, V_m \,\big[
1+ c_m^1\,\epsilon + c_m^2\, \epsilon^2+\cO(\epsilon^3)\big].
\eea
Here $V_p$ and $V_m$ are the field-dependent eigenvalues of 
the matrix $V_{ij}$ of the-tree level $V$:
\bea\label{Vp}
V_p=1/2 \Big[ V_{11}+V_{22}+\big[ (V_{11}-V_{22})^2+4 V_{12}^2\big]^{1/2}\Big]
\eea
with $V_m$ of similar expression but with $-$ in front of the square root.
$V_p$ and $V_m$ should not be confused
with derivatives $V_i$ of the potential.
We also use the notation
\bea
\overline\ln A = \ln\frac{A}{t (z\sigma)^2}-1,
\qquad t=4\pi e^{-\gamma_E} \,.
\eea

\medskip\noindent
Then $V^{(2)}$ and $V^{(2,n)}$  of eq.(\ref{V2}), (\ref{finalU}) 
are shown below. $V^{(2)}$ is that of the
theory regularized with $\mu=$constant, while $V^{(2,n)}$ is a new correction.
They are sums of the  diagrams of eq.(\ref{twoloop})
\bea\label{B5}
V^{(2)} & = & V_{2, old}^a+V_{2, old}^b+V_{2, old}^c
\nonumber\\
V^{(2,\text{n})} & = & V_{2,n}^a+V_{2,n}^b+V_{2,n}^c
\eea
where  $a,b, c,$ label the sunset ($a$), snowman ($b$), counterterm ($c$) diagrams,
respectively.
Then, in terms of the above coefficients, one finds for the sunset diagram ($a$):
\bea
V^a_{2,n} & =&  \!\frac{1}{4 \kappa^2} \, \Big\{
\Big[ v_{ij}w_{lmn}  + \frac{1}{2} u_{ijk}u_{lmn} \Big]\big[ V_p a_{il} + V_m b_{il} \big] 
\big[ a_{jm} + b_{jm} \big] \big[ a_{kn} + b_{kn} \big] 
\nonumber\\
& +&\!\!\! v_{ijk} u_{lmn} \Big[ 
 \Big( V_p\big[ a_{il}(1 \!-\! 2\barlogb \!+\! c_p^1)+ a^1_{il} \big]
 + V_m \big[b_{il}(1 \!-\! 2\barlogb \!+\! c_m^1) + b^1_{il} \big] 
\Big)\big( \!a_{jm}\! +\! b_{jm}\! \big)
\nonumber\\
& +&
 2\big[ V_p a_{il} + V_m b_{il} \big]\big[ a_{jm}^1+b_{jm}^1 \big] \Big]
\big[ a_{kn} + b_{kn} \big] 
+ 
v_{ijk} v_{lmn} \,\,
\Big[
\frac{1}{2}\Big( V_p\big[a_{il}(c_p^2-c_p^1)+a^1_{il}c_p^1 
\nonumber\\
&-& 2\barloga(a_{il}c_p^1 + a_{il}^1) \big] 
% \nonumber\\
 +
 V_m\big[ b_{il}(c_m^2-c_m^1) + b_{il}^1 c_m^1 -2 \barlogb(b_{il}c_m^1 + b_{il}^1) \big] \Big)
\big(a_{jm} + b_{jm} \big)
\nonumber \\
& +&
\Big(V_p a_{il} \big[c_p^1-2\barloga\big] + V_m b_{il}\big[c_m^1-2\barlogb\big] \Big)
\big(a_{jm}^1+b_{jm}^1\big)\Big] \big[ a_{kn} + b_{kn} \big]
\nonumber\\
&+&
 v_{ijk} v_{lmn}  
\Big[
 V_p \Big( b_{il} \big[2 a_{jm}^1 a^1_{kn} + a^2_{jm} b_{kn} + a^1_{jm} b_{kn} + 2 a^1_{jm} b^1_{kn} \big] 
+ a_{il} \big[3a_{jm}^1 a_{kn}^1 + 4a^2_{jn} b_{kn} 
\nonumber\\
&+& 2 b_{jm} b_{kn}^1 + 4 a^1_{jm}(b_{kn} + b^1_{kn}) 
+ 2 b_{jm} b_{kn}^2 \big] 
%\nonumber\\
  + 
 a_{il} a_{jm} \big[3(a_{kn}^1 + a^2_{kn}) + 2(b_{kn}^1 + b^2_{kn}) \big] \Big) 
\nonumber\\
&+ & \!\! V_m \Big( a_{il} \big[2 b_{jm}^1 b^1_{kn} + b^2_{jm} a_{kn} + b^1_{jm} a_{kn} 
+ 2 b^1_{jm} a^1_{kn} \big] 
 +  b_{il} \big[3b_{jm}^1 b_{kn}^1 + 4b^2_{jn} a_{kn} + 2 a_{jm} a_{kn}^1
\nonumber\\
& +& 4 b^1_{jm}(a_{kn}
 + a^1_{kn}) + 2 a_{jm} a_{kn}^2 \big] 
%\nonumber\\
 + 
 b_{il} b_{jm} \big[3(b_{kn}^1 + b^2_{kn}) + 2(a_{kn}^1 + a^2_{kn}) \big] \Big)\,
\Big] \; 
\Big\}
\eea

\vspace{-0.3cm}
\noindent
and
\vspace{-0.2cm}
\bea
V_{2,old}^{a}
&=& 
\frac{1}{4 \kappa^2} \,  v_{ijk} v_{lmn} 
\Big\{ 
 \big[\, V_p a_{il} \barloga + V_m b_{il}\,\barlogb \big] \big[\, a_{jm}a_{kn} 
\barloga + b_{jm}b_{kn} \barlogb \, \big] 
\nonumber\\
&+& 
2 \big[\, v_p a_{il} \overline{\ln}^2 V_p + 
V_m b_{il} \overline{\ln}^2 V_m \big] a_{jm} b_{kn} 
+ \frac{1}{2}
\big[ V_p b_{il} - V_m a_{il} \big] \big[ \overline{\ln}^2 V_p 
- \overline{\ln}^2 V_m\,  \big] a_{jm} b_{kn} 
\nonumber\\
&- & 
\big[V_p a_{il} \barloga + V_m b_{il} \barlogb \big] \big[a_{jm} + 
b_{jm} \big] \big[ a_{kn} + b_{kn} \big] 
\nonumber \\
&+ & \big[V_p a_{il} + V_m b_{il} \big]\big[a_{jm} + b_{jm} \big] \big[ a_{kn}
 + b_{kn} \big] \Big[ \frac{3}{2} + \frac{\pi^2}{12} \Big] 
\nonumber \\
&-& \Big[V_p \big(2a_{il}\Phi_{p,m} - \frac{1}{2} b_{il} \Phi_{m,p} \big) 
+ V_m \big(2b_{il}\Phi_{m,p} - \frac{1}{2} a_{il} \Phi_{p,m} \big) \Big] a_{jm}b_{kn} 
\nonumber\\
&-& \frac{1}{3}\, \big[\, V_p a_{il}a_{jm}a_{kn} + V_m b_{il}b_{jm}b_{kn}\, \big] C
\Big\}, 
\eea
with  \cite{DT}
\bea
\Phi_{p,m} &=&
\begin{cases}
\sqrt{\frac{y_{pm}}{y_{pm}-1}}
\Big[-4\, \mathrm{Li}_2\left( \frac{1-\eta_{pm}}{2}\right) 
+ 2\ln^2 \frac{1-\eta_{pm}}{2} - \ln^2 4y_{pm} + \frac{\pi^2}{3} \Big],& y_{pm}>1
\\[4pt]
4\sqrt{\frac{y_{pm}}{1-y_{pm}}} 
\mathrm{Cl}_2\left(\arcsin \sqrt{y_{pm}}\right),& \!\!\!\!\! 1> y_{pm} \geqslant 0
\end{cases}\;,
\\
y_{pm}& =& {V_m/V_p},
\qquad \eta_{pm} = (1- 1/y_{pm})^{1/2},
\quad C = -2 \sqrt{3} \mathrm{Cl}_2\left( \pi/3 \right)\cong 3.5\,\,\,
\\
\mathrm{Li}_2(\xi) &=&  -\int_0^1 \mathrm{d}t \frac{\ln(1-\xi t)}{t},
\qquad \mathrm{Cl}_2(\theta) = -\int_0^\theta \mathrm{d} \theta \ln 
\left| 2 \sin \frac{\theta}{2} \right|. \,\,
\eea

%%%%%%%%%%%%%%%%%%%%%%%%  

\medskip\noindent
Further, for the snowman diagram ($b$): 
\medskip
\bea
\!\!\! V_{2,n}^b\!\!\!
&=& \!\!\!
\frac{1}{8\,\kappa^2}
\Big\{
w_{ijkl} \,\,(a_{ij} V_p+b_{ij} \,V_m)\,(a_{kl} V_p+b_{kl} \,V_m)\,
-  2\,  
%%% this 2 removed ij<-> kl.
u_{ijkl}\, \Big[
\,
(a_{ij} V_p+b_{ij} \, V_m)
\nonumber\\
&\times&
\Big(\,
a_{kl}\, V_p \,\overline \ln V_p+ b_{kl}\, V_m\, \overline \ln V_m
- 
\big[
(a_{kl} c_{p1}+a_{kl}^1)\,V_p+
(b_{kl} c_{p1}+b_{kl}^1)\,V_m\big]\Big)
%+(ij)\leftrightarrow (kl)
\Big]
\nonumber\\
&-&  2\,
v_{ijkl}
\,\Big[
2\, a_{ij} \,(a_{kl}\, c_{p1} +a_{kl}^1)\, V_p^2 \,\overline \ln V_p
+2 b_{ij}\, (b_{kl}\, c_{m1} +b_{kl}^1)\, V_m^2\,\overline \ln V_m
\nonumber\\
&+&
\big[a_{ij}\, (b_{kl}\,c_{m1}+b_{kl}^1)+ b_{ij}\, (a_{kl}\, c_{p1}+a_{kl}^1)\,\big]
\,V_p\, V_m\,(\overline\ln V_p+\overline\ln V_m)
-
(a_{ij} V_p+b_{ij}\,V_m)\,
\nonumber\\
&\times &
\Big(
\big[ \,a_{kl}\, (c_{p2}-c_{p1}) + a_{kl}^1 c_{p1}+a_{kl}^2\big]\, V_p+
\big[\, b_{kl}\, (c_{m2}-c_{m1}) + b_{kl}^1 c_{m1}+b_{kl}^2\big]\, V_m\Big)
\nonumber\\
&-&\!\!\!\frac{1}{2} 
\big[ ( a_{ij} c_{p1} +a_{ij}^1) V_p+ (b_{ij} c_{m1}+ b_{ij}^1) V_m\big]
\big[ (a_{kl} c_p^1 +a_{kl}^1 ) V_p + (b_{kl} c_{m1}+ b_{kl}^1) V_m \big]
%&+& (ij)\leftrightarrow (kl)
\Big]
\Big\},\qquad
\eea

\medskip\noindent
where $ i,j,k,l=1,2.$ Also
\bea
V_{2,old}^b
& =&
 \frac{1}{8 \kappa^2} v_{ijkl} 
\Big\{ \big[ V_p a_{ij} + V_m b_{ij} \big] 
\big[ V_p a_{kl} + V_m b_{kl} \big] \bigg[ 1+\frac{\pi^2}{6} \bigg]
+ V_p V_m a_{ij} b_{kl} \big[ \barloga - \barlogb \big]^2 
\nonumber\\
&+& 
 2\, \big[\, V_p a_{ij} \barloga + V_m b_{ij} 
\barlogb\, \big]\big[ \, V_p a_{kl} \barloga + V_m b_{kl} \barlogb\, \big]  
 \Big\}.
\eea

\medskip\noindent
For the final ``counter-term'' diagram ($c$)
we need to  introduce the coefficients 
 $\delta v_{ij}$, $\delta u_{ij}$, $\delta w_{ij}$ whose values will be presented shortly
(Appendix C). From eq.(\ref{ZV})
\medskip
\bea\label{co}
\delta V_1 &=& 
\frac{1}{\varepsilon\,\kappa}\,\,\mu^{2\epsilon}
\Big[ \delta^\phi_0\, \frac{\laf}{4!} \phi^4 + 
\delta^m_0 \,\frac{\lam}{4} \phi^2 \sigma^2 
+ \delta^\sigma_0\, \frac{\las}{4!} \sigma^4\Big], \,\,\,\text{then}
\nonumber\\
(\delta V_1)_{ij}  &=& 
\frac{1}{\varepsilon\,\kappa}\,\, \mu^{2\epsilon} \Big[\,
 \delta v_{ij} + \epsilon\, \delta u_{ij}
 + \epsilon^2\, \delta w_{ij}\, \Big],
\eea

\medskip\noindent
where $(\delta V_1)_{ij}=\partial^2(\delta V_1)/\partial s_i \partial s_j$, $i,j=1,2$,
 $s_1,s_2=\phi,\sigma$.
With this notation,  we find for diagram $(c)$:
\bea
V_{2,n}^c   & = &
\frac{1}{2\kappa^2}
\Big\{ \delta w_{ij} \Big[ a_{ij} V_p + b_{ij} V_m \Big] 
+  \delta u_{ij} \Big[ V_p\big( a_{ij}[c_p^1 - \barloga ] + a_{ij}^1 \big)
\nonumber\\
& +&  V_m \big( b_{ij}[c_m^1 - \barlogb ] + b_{ij}^1 \big)\Big] 
-   \delta v_{ij} \Big[
V_p \barloga \big(a_{ij} c_p^1 + a_{ij}^1 \big) + V_m \barlogb \big(b_{ij}c^1_m + b_{ij}^1 \big)  
\nonumber\\
& - &
 V_p \big(a_{ij} c_p^2 + [a^1_{ij} - a_{ij}]c_p^1 + a^2_{ij} \big)
 - V_m \big(b_{ij} c_m^2 + [b^1_{ij} - b_{ij}] c^1_m + b_{ij}^2 \big) 
 \Big]
\Big\}
\eea
and 
\bea
V_{2,old}^c
 = \frac{1}{4 \kappa^2} 
\delta v_{ij}\Big[ V_p a_{ij}\Big(\, \overline{\ln}^2 V_p + 1+\frac{\pi^2}{6} \Big)
 +  V_m b_{ij}\Big( \overline{\ln}^2 V_m + 1+\frac{\pi^2}{6} \,\Big) \Big].
\eea

 \def\theequation{C-\arabic{equation}}\def\thesubsection{C}
\subsection{Coefficients entering the two-loop potential}
\setcounter{equation}{0}
 \label{appendixC}

\noindent
Below we provide the expressions of the various 
coefficients introduced in the rhs of eq.(\ref{intro}), (\ref{co}) and
used in  Appendix~\ref{appendixB}.
The coefficients  $v_{ijkl}$, $v_{ijk}$ are functions
of $\lambda$'s and fields
\medskip
\begin{gather}
\left[
\begin{array}{ccc}
v_{1111} & u_{1111} & w_{1111} \\
v_{1112} & u_{1112} & w_{1112} \\
v_{1122} & u_{1122} & w_{1122} \\
v_{1222} & u_{1222} & w_{1222} \\
v_{2222} & u_{2222} & w_{2222} \\
\end{array}
\right]
\!=\!
\left[
\begin{array}{ccc}
 \lambda _{\phi } & 0 & 0 \\
 0 & 2\laf \frac{\phi}{\sigma } & 2\laf \frac{ \phi}{\sigma } \\
 \lambda _m & 3 \lambda _m- \laf \frac{\phi ^2 }{\sigma ^2} &
 \laf\frac{\phi ^2 }{\sigma ^2}+5 \lambda _m \\
 0 & \frac{2}{3}\laf\frac{\phi ^3 }{\sigma ^3}+
 2\lam\frac{ \phi }{\sigma } & 8 \lam \frac{ \phi  }{\sigma }
- \frac{4}{3} \laf\frac{\phi ^3}{ \sigma ^3} \\
 \lambda _{\sigma } & -\frac{1}{2}\laf \frac{\phi ^4}{\sigma ^4}
+\frac{25}{6}\las-\lam\frac{\phi ^2}{\sigma ^2} &
  \frac{4}{3} \laf \frac{\phi ^4}{\sigma ^4}+10 \lambda _{\sigma }- 2\lam\frac{ \phi ^2 }{\sigma ^2} \\
\end{array}
\right]
\end{gather}
and 
\medskip
\begin{gather}
\left[
\begin{array}{ccc}
v_{111} & u_{111} & w_{111} \\
v_{112} & u_{112} & w_{112} \\
v_{122} & u_{122} & w_{122} \\
v_{222} & u_{222} & w_{222} \\
\end{array}
\right]
\!=\!
\left[
\begin{array}{ccc}
 \laf \phi   & 0 & 0 \\
 \lam \sigma & \laf\frac{\phi ^2}{\sigma }+\sigma  \lambda _m & \laf 
\frac{\phi ^2 }{\sigma }+ \lam \sigma  \\
 \lam \phi  & 3 \lam \phi -\frac{1}{3} \laf\frac{\phi ^3}{\sigma ^2} &
 \frac{1}{3}\laf\frac{\phi ^3 }{ \sigma ^2}+5 \lam \phi 
   \\
 \las \sigma  & \frac{1}{6} \laf \frac{\phi ^4 }{\sigma ^3}+\frac{13}{6}
 \las \sigma  + \lam \frac{\phi ^2 }{\sigma } & -\frac{1}{3}\laf \frac{\phi ^4 }{ \sigma ^3}
+\frac{11}{3} \las \sigma +4 \lam \frac{ \phi ^2}{\sigma }
   \\
\end{array}
\right].
\end{gather}

\medskip\noindent
Further, coefficients $\delta v_{ij}$, $\delta u_{ij}$ and $\delta w_{ij}$ of eq.(\ref{co}) are
\medskip
\bea
\delta v_{11}  \!\!&=&\!\!
 \frac{1}{4\kappa} \left[
3\left(\laf^2 + \lam^2 \right) \phi^2 + \lam \left(\laf + 4\lam + \las \right) \sigma^2 
 \right]
   \;, \quad
   \delta u_{11} = \delta w_{11} = 0
\\
\delta v_{12} \!\!&=&\!\! \frac{1}{2\kappa} \lam \left( \laf + 4 \lam + \las \right)\phi \sigma \\
\delta u_{12} \!\!&=&\!\! \delta w_{12} = \frac{1}{2 \kappa} \left[ \left( \laf^2 + \lam^2 \right) \phi^2 +
\lam \left(\laf + 4 \lam + \las \right) \sigma^2
\right] \frac{\phi}{\sigma}
\\
\delta v_{22} \!\!&=&\!\! \frac{1}{4\kappa^2} \left[
\left(4 \lam^2 + \laf \lam + \lam \las \right) \phi^2 + 
3\left(\lam^2 + \las^2 \right)\sigma^2 \right]
\\
\delta u_{22} \!\!&=&\!\!
\frac{1}{8 \kappa^2} \frac{1}{\sigma^2}\left[
-\left( \laf^2 + \lam^2 \right)\phi^4 + 
6\lam\left( \laf + 4\lam + \las \right)\phi^2 \sigma^2 +
7\left( \las^2 + \lam^2 \right) \sigma^4
\right]
%\nonumber
\\
\delta w_{22} \!\!&=&\!\!
\frac{1}{8 \kappa^2} \frac{1}{\sigma^2}\left[
\left( \laf^2 + \lam^2 \right)\phi^4 + 
10\lam\left( \laf + 4\lam + \las \right)\phi^2 \sigma^2 +
9\left( \las^2 + \lam^2 \right) \sigma^4
\right]\qquad
\eea

\bigskip\noindent
The
coefficients $c_p^1$, $c_p^2$, $c_m^1$, $c_m^2$, also $V_p$ and $V_m$  introduced in
eq.(\ref{intro}) have  the expressions:
\medskip
\bea
\left[
\begin{array}{c}
V_p\, c_p^1 \\
V_m\, c_m^1  \\
\end{array}
\right]
\!=\!
\frac{1}{24}\frac{1}{\sigma^2}
\left[
\begin{array}{c}
-\laf \phi ^4 + 18 \lam \phi ^2\sigma^2+7 \las \sigma ^4 + R_1 \\
-\laf \phi ^4 + 18 \lam \phi ^2\sigma^2+7 \las \sigma ^4 - R_1 \\
\end{array}
\right]
\eea

\medskip\noindent
where
\bea
R_1& =& \frac{1}{2 S} \Big[
\laf\left(\laf - \lam \right)\phi^6 +
\left( 15 \laf \lam - \laf \las + 18 \lam^2 \right)\phi^4 \sigma^2 
\nonumber\\
&+&
\left( -7\laf\las + 78\lam^2 + 25\lam\las \right)\phi^2 \sigma^4 +
7\las\left( -\lam + \las \right)\sigma^6
   \Big]
\eea
and 
\medskip
\be\label{S}
S^2 = \frac{1}{4} \Big[
\left( \laf - \lam \right)^2 \phi^4 +
2\left( \laf \lam + 7 \lam^2 - \laf \las + \lam \las \right) \phi^2 \sigma^2 + 
\left(\lam -\las \right)^2 \sigma^4
\Big],
\ee

\medskip\noindent
while $V_m$ and $V_p$ (see also (\ref{Vp})) are
\medskip
\bea
\left[
\begin{array}{c}
V_p  \\
V_m  \\
\end{array}
\right]
\!=\!
\frac{1}{4}
\left[
\begin{array}{c}
 \left(\lambda _{\sigma }+\lambda _m\right) \sigma ^2  
+ \left(\lambda _{\phi }+\lambda _m\right)\phi ^2 + 2S \\
 \left(\lambda _{\sigma }+\lambda _m\right) \sigma ^2  
+ \left(\lambda _{\phi }+\lambda _m\right)\phi ^2 - 2S 
\end{array}
\right].
\eea
For $c_p^2$, $c_m^2$ (with the above $V_p$ and $V_m$):
\bea
\left[
\begin{array}{c}
V_p \,  c_p^2 \\
V_m \, c_m^2  \\
\end{array}
\right]
\!=\!
\frac{1}{24}\frac{1}{\sigma^2}
\left[
\begin{array}{c}
\laf \phi ^4 + 30 \lam \phi ^2\sigma^2 + 9 \las \sigma ^4 + R_2\\
\laf \phi ^4 + 30 \lam \phi ^2\sigma^2 + 9 \las \sigma ^4 - R_2
\end{array}
\right],
\eea
where
%\medskip
\bea
\!\! R_2 &=& \!\!\frac{1}{\sigma^2 S^3}
\Big[
\sigma ^2 \phi ^{10} \lambda _{\phi } \big(13 \lambda _{\phi }^3
+27 \lambda _{\phi } \lambda _m^2-39 \lambda _{\phi }^2 \lambda _m+3 \lambda _m^3\big)
  +  \sigma ^4 \phi ^8 \big(-23 \lambda _{\sigma } \lambda _{\phi }^3
\nonumber\\
& + &
3 \lambda _{\phi }  \lambda _m^2 \left[3 \lambda _{\sigma }+16 \lambda _{\phi }\right]
+5  \lambda _{\phi }^2 \lambda _m \left[6 \lambda _{\sigma }
+25 \lambda _{\phi }\right]-423\lambda _{\phi } \lambda _m^3+90 \lambda _m^4\big)
\nonumber\\
&+&
\sigma ^6 \phi ^6 \big(\lambda _{\sigma } \lambda _{\phi }^2 
\left[7 \lambda _{\sigma }-27 \lambda _{\phi}\right]+3 \lambda _m^3 \left[99 \lambda _{\sigma }+151 \lambda _{\phi }\right]
+\lambda _{\phi } \lambda _m^2 \left[553 \lambda _{\phi }-
615 \lambda _{\sigma }\right]
\nonumber\\
&+ & 
\lambda _{\sigma } \lambda _{\phi } \lambda _m 
\left[9 \lambda _{\sigma }+65\lambda _{\phi }\right]+2034 \lambda _m^4\big)
%\nonumber\\
+
\sigma ^8 \phi ^4 \big(3 \lambda _{\sigma }^2 
\lambda _{\phi } \left[\lambda _{\sigma }+27 \lambda _{\phi }\right]
+3 \lambda _m^3 \left[641 \lambda _{\sigma } + 261 \lambda _{\phi }\right]
\nonumber\\
&+ & 
\lambda _{\sigma } \lambda _m^2 \left[351 \lambda _{\sigma }
-521\lambda _{\phi }\right]-\lambda_{\sigma } \lambda _{\phi } 
\lambda _m \left[361 \lambda_{\sigma }+81 \lambda _{\phi }\right]+3438 \lambda_m^4\big)
+
\sigma ^{10} \phi ^2 \big(-81 \lambda _{\sigma }^3 \lambda_{\phi }
\nonumber\\
&+&
\lambda _{\sigma } \lambda _m^2 \left[292 \lambda _{\sigma }-81\lambda _{\phi }\right]
+
9 \lambda _{\sigma}^2 \lambda _m \left[19 \lambda _{\sigma }
+18 \lambda _{\phi }\right]-609 \lambda _{\sigma } \lambda _m^3+342 \lambda _m^4\big)
\nonumber\\
&-&
27 \sigma ^{12} \lambda _{\sigma } \left[\lambda _m-\lambda _{\sigma }\right]^3
   \Big].
\eea

\medskip\noindent
Finally, 
 $a_{ij}$, $a_{ij}^1$, $a_{ij}^2$, $b_{ij}$, $b_{ij}^1$, $b_{ij}^2$
introduced  in (\ref{intro}) and used in Appendix~\ref{appendixB} have the values
\medskip
\bea
a_{11}& =& 1-a_{22} = b_{22}=1 - b_{11} = 
\frac{1}{2} + \frac{1}{4S}\big[\laf \phi^2 + \lam(-\phi^2 + \sigma^2) - \las \sigma^2 \big]
\\
a_{12}& =& a_{21} = - b_{12} = - b_{21} =\frac{\lam \phi\, \sigma}{S} 
\\
a_{11}^1 &=& \!\! -a_{22}^1 = - b_{11}^1 = b_{22}^1 
=\frac{\lam \phi^2}{6\, S^3}
\Big[
\laf \big(-2\laf + 3 \lam \big) \phi^4 
\nonumber\\
&& \hspace{2cm}+\,  2\big(\laf\las -4\laf\lam 
-
 6 \lam^2 \big) 
\phi^2 \sigma^2 -\big(6\lam^2 + \lam \las \big) \sigma^4
\Big]
\eea
with $S$ of eq.(\ref{S}).
Also
\bea
a_{12}^1 &=& a_{21}^1 = -b_{12}^1 = - b_{21}^1 = \frac{\phi}{24 \sigma S^3} \Big[
 \phi ^6 
   \lambda _{\phi } \left(2 \lambda _{\phi }^2-5 \lambda
   _{\phi } \lambda _m+3 \lambda _m^2\right)\qquad\qquad
\nonumber\\
   &+&
 \phi ^4 \sigma ^2
   \left(-4 \lambda _{\sigma } \lambda _{\phi }^2+5 \lambda _{\phi } \lambda _m \left[\lambda _{\sigma }+2
   \lambda _{\phi }\right]+\lambda _{\phi } \lambda _m^2-12 \lambda _m^3\right)
\nonumber   \\
   &+&
 \phi ^2 \sigma ^4
   \left(2 \lambda _{\sigma }^2 \lambda _{\phi }+\lambda _m^2 \left[14
 \lambda _{\phi }-13 \lambda_{\sigma }\right]-9 \lambda _{\sigma } \lambda _{\phi } \lambda _m+6 \lambda _m^3\right) 
\nonumber   \\
   &+&
 \sigma ^6
  \lambda _m \left(-\lambda _{\sigma }^2-5 \lambda _{\sigma } \lambda_m+6 \lambda _m^2\right)
 \Big].
 \eea
 Further
 \bea
 a_{11}^2  &=& -a_{22}^2=-b_{11}^2 = b_{22}^2 
=
  \frac{\phi^2}{288 \sigma^2 S^5} 
\Big\{     \phi ^{10}\,     \lambda _{\phi }^2 \,
\big(-4 \lambda _{\phi }^3
-31 \lambda _{\phi } \lambda _m^2
+20 \lambda _{\phi }^2 \lambda _m+15 \lambda _m^3\big)
\nonumber   \\
   &+&
\!\! \phi ^8 \sigma ^2  \lambda _{\phi }\, \Big[
12 \lambda _{\sigma } \lambda _{\phi }^3
+\lambda _{\phi } \lambda _m^2 \left[31 \lambda _{\sigma }+180 \lambda _{\phi }\right]
%\nonumber   \\
- 
20 \lambda _{\phi }^2 \lambda _m \,\left[2 \lambda _{\sigma }
+3 \lambda _{\phi }\right] +5 \lambda _{\phi } \lambda _m^3-192 \lambda
   _m^4\Big]
\nonumber   \\
  & +&
2   \phi ^6 \sigma ^4
  \Big[ -6 \lambda _{\sigma }^2 \lambda _{\phi }^3
+\lambda _{\phi } \lambda _m^3 \left[44 \lambda _{\phi }
-163 \lambda _{\sigma }\right]-\lambda _{\phi }^2 \lambda _m^2 
\left[19 \lambda _{\sigma }+132 \lambda _{\phi }\right]
 \nonumber  \\
& + & 
2 \lambda _{\sigma } \lambda _{\phi }^2 \lambda _m
 \left[5 \lambda _{\sigma }+28 \lambda _{\phi }\right]+450
   \lambda _{\phi } \lambda _m^4+144 \lambda _m^5\Big]
%\nonumber  \\
 -
 2 \phi ^4 \sigma ^6 
  \Big[ -2 \lambda _{\sigma }^3 \lambda _{\phi }^2
+\lambda _m^4 \left[270 \lambda _{\phi }
\right.\nonumber\\
& -& \left.
96 \lambda _{\sigma }\right]+\lambda _{\phi } \lambda _m^3
 \left[236 \lambda _{\phi }-463 \lambda _{\sigma }\right]
+
\lambda _{\sigma } \lambda _{\phi } \lambda _m^2 
\left[71 \lambda _{\sigma }-118 \lambda _{\phi }\right]+22
   \lambda _{\sigma }^2 \lambda _{\phi }^2 \lambda _m+1008 \lambda _m^5\Big]
\nonumber  \\
 &+&
  \phi ^2 \sigma ^8
  \lambda _m\, \Big[
-8 \lambda _{\sigma }^3\lambda _{\phi }+12 \lambda _m^3
 \left[29 \lambda _{\sigma }-31 \lambda_{\phi }\right]
% \nonumber \\
+
\lambda _{\sigma } \lambda _m^2 \left[184 \lambda _{\phi}
-117 \lambda _{\sigma }\right]+49 \lambda _{\sigma }^2 \lambda _{\phi }\lambda _m
\nonumber\\
& -&  468 \lambda _m^4\Big]
 + \sigma ^{10}
 3\lambda _m^2 \big(-7 \lambda _{\sigma }^3+16 \lambda _{\sigma} \lambda _m^2
+27 \lambda _{\sigma }^2 \lambda _m-36 \lambda_m^3\big)
  \Big\}.
 \eea
 Finally
 \bea
 a_{12}^2 &=& a_{21}^2 = -b_{12}^2 = -b_{21}^2 
=
 -
\frac{\phi}{576\sigma^3 S^5} \Big[
   - \phi ^{12} \lambda _{\phi }^2 \left(3  \lambda _m
-2 \lambda _{\phi }\right) \big(\lambda _m-\lambda _{\phi }\big)^2
 \nonumber\\
   &+&
   2 \sigma ^2 \phi ^{10} \lambda _{\phi } \big(
-3 \lambda _{\phi }^3 \left[\lambda _{\sigma }+
2 \lambda _{\phi }\right]-\lambda _{\phi } \lambda _m^2 \left[4 \lambda_{\sigma } 
 +99 \lambda _{\phi }\right]
+\lambda _{\phi }^2 \lambda _m  \left[7 \lambda _{\sigma }
+57 \lambda _{\phi }\right]
\nonumber\\
&+ & 
22 \lambda _{\phi }  \lambda _m^3+30 \lambda _m^4\big)  
+
      \sigma ^4 \phi ^8 \big(2 \lambda _{\sigma } \lambda _{\phi  }^3 
\left[3 \lambda _{\sigma }+17 \lambda _{\phi }\right]+\lambda _{\phi } 
 \lambda _m^3 \left[160 \lambda _{\sigma }+301 \lambda _{\phi }\right]
\nonumber\\
& +& 
2 \lambda _{\phi }^2 \lambda _m^2 \left[28 \lambda _{\sigma }+237 \lambda
   _{\phi }\right]
-\lambda _{\phi }^2 \lambda _m
 \left[184 \lambda _{\sigma }  \lambda _{\phi }+7 \lambda _{\sigma }^2
+84 \lambda _{\phi }^2\right]-972  \lambda _{\phi } \lambda _m^4-72 \lambda _m^5\big)
 \nonumber\\
 &+&
   2 \sigma ^6 \phi ^6 \big(-\lambda_{\sigma }^2 \lambda _{\phi }^2 
\left[\lambda _{\sigma }+15 \lambda_{\phi }\right]-6 \lambda _m^4 \left[7 \lambda _{\sigma }
-68 \lambda _{\phi }\right] +\lambda _{\phi } \lambda _m^3 \left[379 \lambda _{\phi }
-651 \lambda _{\sigma }\right]
 \nonumber\\
&&+\lambda _{\phi } \lambda _m^2 \left[-5 \lambda _{\sigma } 
\lambda _{\phi }+71 \lambda _{\sigma }^2-108 \lambda _{\phi}^2\right]
+\lambda _{\sigma } \lambda _{\phi }^2 \lambda _m \left[13
 \lambda _{\sigma }+75 \lambda _{\phi }\right]+1116 \lambda _m^5\big)
\nonumber \\
   &+&
   \sigma ^8 \phi ^4 \big(6 \lambda _{\sigma }^3 \lambda _{\phi }^2+12 
\lambda _m^4 \left[92 \lambda _{\sigma }+45 \lambda _{\phi }\right]
+\lambda _m^3 \left[540 \lambda_{\sigma } \lambda _{\phi }
+59 \lambda _{\sigma }^2-264 \lambda _{\phi }^2\right]
 \nonumber \\
&&-6 \lambda _{\sigma } \lambda _{\phi } \lambda _m^2  \left[91
 \lambda _{\sigma }-41 \lambda _{\phi }\right]+\lambda _{\sigma }^2 \lambda _{\phi } \lambda _m \left[44
 \lambda _{\sigma }-37 \lambda _{\phi }\right]+324 \lambda _m^5\big)
\nonumber  \\
  &+&
2 \sigma ^{10} \phi ^2 \big(\lambda _{\sigma}^4 \lambda _{\phi }
+6\lambda _m^4 \left[31 \lambda _{\sigma }-13 \lambda _{\phi }\right]
+\lambda _{\sigma } \lambda _m^3 \left[89 \lambda _{\phi
   }-103 \lambda _{\sigma }\right]
\nonumber   \\&&+\lambda _{\sigma }^2 \lambda _m^2 \left[41 
\lambda _{\sigma }+8 \lambda _{\phi }\right]-20 \lambda _{\sigma }^3 \lambda _{\phi } \lambda _m
+72 \lambda _m^5\big) 
\nonumber   \\ 
   &+&
\sigma ^{12} \left(-\lambda _m\right) \left(\lambda _m
-\lambda_{\sigma}\right)^2 \left(-11 \lambda _{\sigma }^2
+24 \lambda _{\sigma } \lambda _m+ 36 \lambda _m^2\right)
  \Big],
\eea

\medskip\noindent
which enter in the expression of the two-loop potential.

\bigskip\bigskip
\noindent
{\bf Acknowledgements:    }
%
% \noindent
The  work of D. Ghilencea  was supported by a grant from Romanian National  Authority for
Scientific Research (CNCS-UEFISCDI) under  project number PN-II-ID-PCE-2011-3-0607.
The work of P. Olszewski  and Z. Lalak was  supported by the Polish NCN grants DEC-2012/04/A/ST2/00099
and 2014/13/N/ST2/02712.

%\newpage

\end{document}